\documentclass[fleqn]{aa}  

\usepackage{graphicx}
\usepackage{xcolor}
\usepackage{hyperref}
\usepackage{txfonts}
\usepackage{calligra}
\usepackage{calrsfs}
\usepackage[T1]{fontenc}
\usepackage[mathcal]{eucal}
\usepackage{longtable}
\newcommand{\vectX}{\bf {\it X}}
\newcommand{\vectTheta}{\bf {\it \Theta}}

\newcommand{\matrGamma}{\bf \Gamma}
\newcommand{\Msun}{\, {\rm M}_{\odot}}
\newcommand{\Lsun}{\, {\rm L}_{\odot}}
\newcommand{\K}{\, {\rm K}}

\begin{document}

   \title{BUFFALO/Flashlights: Constraints on the abundance of lensed supergiant stars in the Spock galaxy at redshift 1.}
   
   \titlerunning{Supergiant stars at redshift 1}
  \authorrunning{Diego et al.}

   \author{Jose M. Diego
         \inst{1}\fnmsep\thanks{jdiego@ifca.unican.es}
       \and
       Sung Kei Li \inst{2} 
        \and  
       Ashish K. Meena\inst{3}
        \and
        Anna Niemiec\inst{4}
       \and
        Ana Acebron\inst{5,6}
       \and
       Mathilde Jauzac\inst{7,8,9,10}
       \and
       Mitchell F. Struble\inst{11}
       \and
Alfred Amruth  \inst{2}
       \and
Tom J. Broadhurst \inst{13,14,15}
       \and
Catherine Cerny \inst{7,8}
    \and
    Harald Ebeling \inst{16}
    \and
Alexei V. Filippenko \inst{16}
       \and
Eric Jullo \inst{18}
       \and
Patrick Kelly \inst{19}
       \and
Anton M. Koekemoer \inst{20}
       \and
David Lagatutta \inst{7,8}
       \and
Jeremy Lim \inst{2}
       \and
Marceau Limousin \inst{18}
       \and
Guillaume Mahler \inst{7,8}
       \and
Nency Patel \inst{8}
    \and
Juan Remolina \inst{21}
       \and
Johan Richard \inst{22}
       \and
Keren Sharon \inst{21}
       \and
Charles Steinhardt \inst{23}
       \and
Keichii Umetsu \inst{24}
       \and
Liliya Williams \inst{25}
       \and
Adi Zitrin \inst{3}
       \and
       J.M. Palencia \inst{1}
       \and
        Liang Dai \inst{26}
        \and 
        Lingyuan Ji \inst{26}
        \and
       Massimo Pascale \inst{17}
    }      
   \institute{Instituto de F\'isica de Cantabria (CSIC-UC). Avda. Los Castros s/n. 39005 Santander, Spain 
        \and
        Department of Physics, The University of Hong Kong, Pokfulam Road, Hong Kong  
        \and
         Physics Department, Ben-Gurion University of the Negev, P.O. Box 653, Be’er-Sheva 84105, Israel 
         \and 
         LPNHE, CNRS/IN2P3, Sorbonne Université, Université Paris-Cité, Laboratoire de Physique Nucléaire et de Hautes Énergies, 75005 Paris, France 
          \and 
         Department of Physics of the University of Milano, via Celoria 16, I-20133 Milano, Italy 
         \and 
         INAF – IASF Milano, via A. Corti 12, I-20133 Milano, Italy 
         \and         
         Centre for Extragalactic Astronomy, Durham University, South Rd, Durham, DH1 3LE, UK 
         \and 
        Institute for Computational Cosmology, Durham University, South Road, Durham DH1 3LE, UK 
         \and   
        Astrophysics Research Centre, University of KwaZulu-Natal, Westville Campus, Durban 4041, South Africa 
         \and 
School of Mathematics, Statistics \& Computer Science, University of KwaZulu-Natal, Westville Campus, Durban 4041, S. Africa 
         \and
         Department of Physics and Astronomy, University of Pennsylvania, 209 South 33rd Street, Philadelphia, PA 19104, USA 
         \and
         Department of Physics, University of Hong Kong, Hong Kong, Hong Kong SAR 
       \and
          Department of Physics, University of the Basque Country UPV/EHU, E-48080 Bilbao, Spain 
         \and
          DIPC, Basque Country UPV/EHU, E-48080 San Sebastian, Spain 
          \and
          Ikerbasque, Basque Foundation for Science, E-48011 Bilbao, Spain 
          \and          
          Institute for Astronomy, University of Hawaii, 640 N Aohoku Pl, Hilo, HI 96720, USA 
         \and
         Department of Astronomy, University of California, Berkeley, CA 94720-3411, USA 
         \and
         Aix Marseille Univ, CNRS, CNES, LAM, F-13388 Marseille, France 
         \and
         Minnesota Institute for Astrophysics, University of Minnesota, 116 Church Street SE, Minneapolis, MN 55455, USA 
         \and
         Space Telescope Science Institute, 3700 San Martin Dr., Baltimore, MD 21218, USA 
         \and
        Department of Astronomy, University of Michigan, 1085 S. University Ave, Ann Arbor, MI 48109, USA 
         \and
          Univ. Lyon, Ens de Lyon, CNRS, Centre de Recherche Astrophysique de Lyon UMR5574, 69230 Saint-Genis-Laval, France 
         \and
        Niels Bohr Institute, University of Copenhagen, Lyngbyvej 2, København Ø 2100, Denmark 
        \and
        Institute of Astronomy and Astrophysics, Academia Sinica (ASIAA), AS/NTU Astronomy-Mathematics Building, No. 1, Sec. 4, Roosevelt Rd., Taipei 10617, Taiwan 
        \and
        School of Physics and Astronomy, University of Minnesota, 116 Church Street, Minneapolis, MN 55455, USA 
        \and
        Department of Physics, University of California, 366 Physics North MC 7300, Berkeley, CA 94720, USA 
          }

 \abstract{
    In this work we present a constraint on the abundance of supergiant (SG) stars at redshift $z\approx 1$, based on recent observations of a strongly lensed arc at this redshift. First we derive a free-form model of MACS J0416.1-2403 using data from the BUFFALO program. The new lens model is based on 72 multiply lensed galaxies that produce 214 multiple images, making it the largest sample of spectroscopically confirmed lensed galaxies on this cluster. The larger coverage in BUFFALO allows us to measure the shear up to the outskirts of the cluster, and extend the range of lensing constraints up to $\sim 1$ Mpc from the central region, providing a mass estimate up to this radius. As an application,  we make predictions for the number of high-redshift multiply-lensed galaxies detected in future observations with {\it JWST}. Then we focus on a previously known lensed galaxy at $z=1.0054$, nicknamed Spock, which contains four previously reported transients. We interpret these transients as microcaustic crossings of SG stars and compute the probability of such events. Based on simplifications regarding the stellar evolution, we find that microlensing (by stars in the intracluster medium) of SG stars at $z=1.0054$ can fully explain these events. 
    The inferred abundance of SG stars is consistent with either (1) a number density of stars with bolometric luminosities beyond the  Humphreys-Davidson (HD) limit ($L_{\rm max}\approx 6\times10^5\,\Lsun$ for red stars) that is below $\sim 400$ stars\,kpc$^{-2}$, or (2) the absence of stars beyond the HD limit but with a SG number density of $\sim 9000$\,kpc$^{-2}$ for stars with  luminosities between $10^5\Lsun$ and $6\times10^5\Lsun$. This is equivalent to one SG star per $10\times10$ pc$^2$.
    We finally make predictions for future observations with {\it JWST}'s NIRcam. We find that in observations made with the F200W filter that reach 29 mag AB, if cool red SG stars exist at $z\approx 1$ beyond the HD limit, they should be easily detected in this arc. 
   }

   \maketitle
%

\section{Introduction}
The galaxy cluster MACS J0416.1-2403 \citep[MACS0416 hereafter][]{Mann2012} at redshift $z=0.4$ is one of the six clusters studied with exquisite detail by the {\it Hubble Space Telescope (HST)} as part of the Hubble Frontier Field (HFF) program. It was selected based on its prominent lensing signature \citep{Zitrin2013}. During this program, {\it HST} observed $\sim 10$ arcmin$^2$ around the central portion of the cluster in wavelengths ranging from 0.45 $\mu$m to 1.6 $mu$m and to a depth of $\sim 28.5$ mag in the optical and infrared (IR) bands. The HFF data from MACS01416 have been extensively used by lens modellers \citep[][]{Jauzac2014,Jauzac2015,Johnson2014, Diego2015b, Kawamata2016, Hoag2016, Caminha2017, Richard2021,Bergamini2022} in the search for high-redshift galaxies \citep{McLeod2015,Oesch2015,Kawamata2016,Ishigaki2018}, and for the discovery of some of the first lensed stars beyond $z=0.9$ \citep{Rodney2018,Chen2019,Kaurov2019}. 

One of the goals of the HFF program was to find the most distant galaxies taking advantage of the magnifying boost given by the gravitational lensing effect. The frequency coverage of {\it HST} allows one to potentially detect galaxies up to $z\approx 12$ thanks to the dropout technique \citep{Steidel1999}. The census of these high-redshift galaxies has provided useful information about the evolution of the ultraviolet (UV) luminosity function with redshift. However, the number of galaxies at the highest redshifts remains low, resulting in large uncertainties in the luminosity function. In order to increase the statistics, the BUFFALO program doubles the area around these clusters that is being covered by the IR detectors. In addition, the extended area around the central region of the clusters allows us to complement the strong-lensing constraints with weak-lensing measurements. The addition of weak lensing in the external parts of the cluster is important because it helps to reduce the uncertainty in the magnification, vital to properly derive the luminosity functions of distant lensed galaxies observed around this cluster. 

In the first part of this work we present a new lens model derived with an updated set of strong-lensing constraints, containing new spectroscopically identified lensed galaxies, and combined with new measurements of the shear over the entire BUFFALO field of view.  
As a simple application, during the first part of the work we also use this lens model to predict the number of galaxies at $z=6$ and $z=9$. 

In the second part of this work we focus on a small but highly magnified region in the image plane, the Spock arc \citep{Rodney2018},  where earlier work reports four transient events \citep{Rodney2018,Kelly2023}. This arc is very stretched, containing a highly magnified portion of a lensed galaxy at $z=1.0054$.  Two of the transient events where found as part of the FrontierSN program (HST-PID 13386, PI S. Rodney), and the other two as part of the Flashlights program \citep{Kelly2023}. 
The FrontierSN program is designed to produce light curves of supernovae (SNe), so it has a natural good cadence of $\sim 1$ observation per day during a period of $\sim 1$ month. The Flashlights program is designed to find flux variations of unresolved objects, and to accomplish this, it makes two observations with the same position angle (so subtracted images minimise instrumental effects) using very wide filters (to reach fainter objects). 
The transients in the Spock arc have been interpreted as possible microlensing events of distant stars in the galaxy at $z=1.0054$, which are temporarily being magnified by stars in the cluster (at $z=0.396$). Other possible interpretations for the transients are considered by \cite{Rodney2018} [for instance luminous blue variable (LBV) stars or recurrent novae], but in this work we adopt the hypothesis that the transient nature of these events is due to changes in the magnification produced by microlensing. This choice is motivated by the fact that we do not know if these stars are intrinsically variable (LBV or nova), although many luminous stars are, but we do know that microlenses in the lens plane are relatively numerous in this portion of the lens (from the observed intracluster light), so microlensing events are expected. 

\begin{figure} 
   \includegraphics[width=9cm]{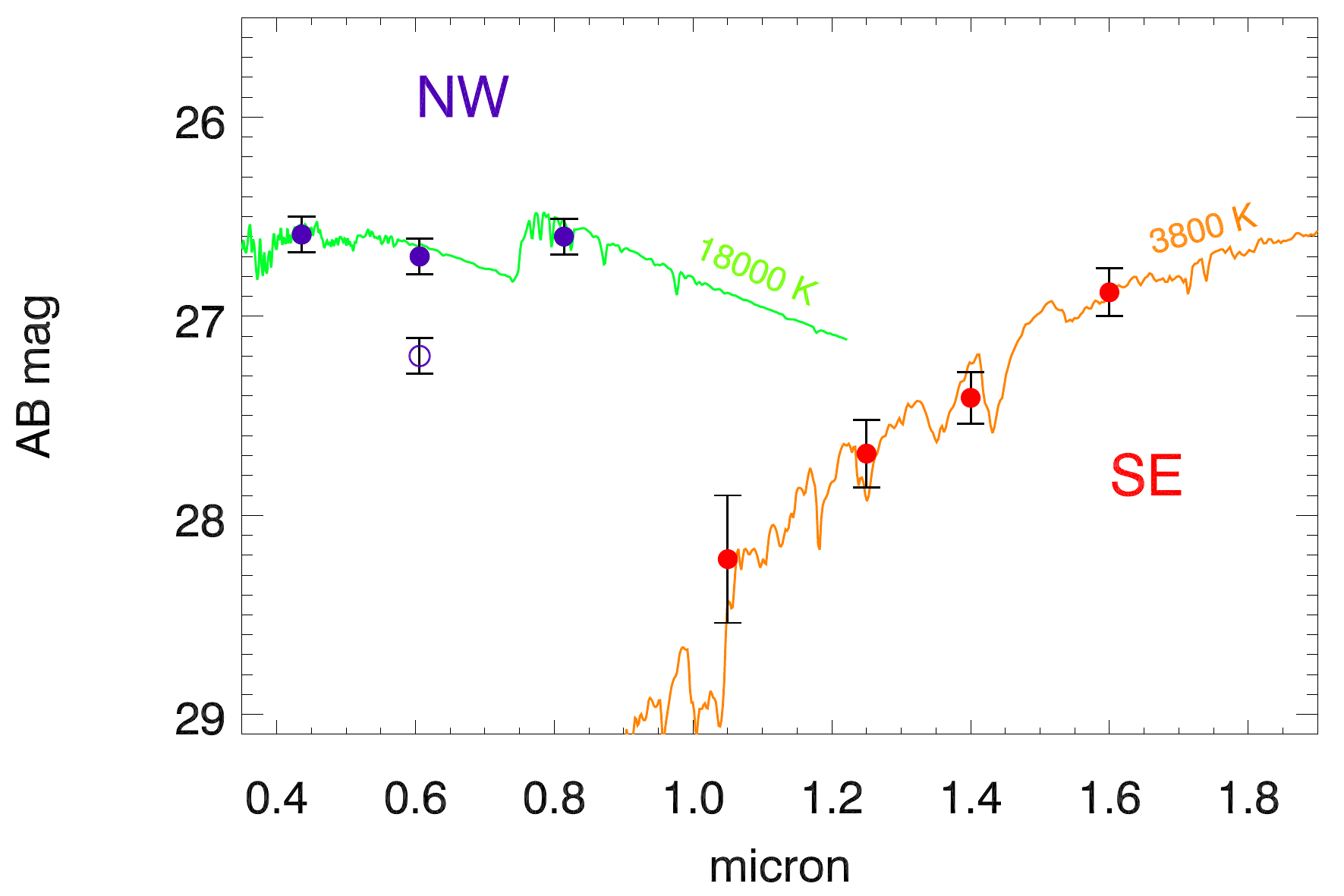} 
      \caption{Photometry of the two events (SE and NW) in the Spock galaxy reported by \cite{Rodney2018} vs. stellar models. For each event, the magnitudes in this plot correspond to the moment of maximum observed luminosity in \cite{Rodney2018} and are obtained from that work. The colored curves are synthetic models from \cite{Coelho2014} for different star temperatures. The SE event is well described by a red SG with $T\approx3,800\K$, while the NW event is better described by a hotter star system with temperature $T\approx 18,000\K$. Blue dots correspond to the optical bands while red dots are for the IR channels. The open circle is the magnitude measured in F606W 1 day (half day in the rest frame) before maximum. We correct this magnitude by 0.5 magnitudes from the relation $dm/dt \approx 1\, mag\, day^{-1}$ found in \cite{Rodney2018} where time is expressed in the rest frame of the Spock galaxy (see their Supplementary Figure 9).
      }
         \label{Fig_Photometry}
\end{figure}

As shown in Figure~\ref{Fig_Photometry}, the observed photometry during the maximum observed luminosity is also consistent with the hypothesis that the two events of \cite{Rodney2018} are very luminous stars undergoing microlensing episodes near the critical curve of the cluster. In particular, we find that one of the events is well described by a red supergiant (SG) star, while the other one is better described by a hotter SG star. Data points shown in this figure correspond to the measured photometry around the position of maximum luminosity. The photometry in the F606W filter (0.6 micron) has been corrected to account for the fact that this data point corresponds to 1 day (half day in the rest frame) before maximum luminosity. The two events from the Flashlights program were obtained with a wide filter, so we do not have color information that can be used to test the star hypothesis by comparing the photometry with stellar models.

Under the microlensing hypothesis, the stars that one expects to detect are very luminous. Among these, hot blue stars  can have luminosities exceeding $10^6\,\Lsun$. Red (cooler) stars are typically less luminous.  
Local measurements show that the most luminous red SG stars, such as UY Scuti or Stephenson 2-18, have luminosities below $6\times10^5\,L_{\odot}$. Similar results are found in nearby galaxies \citep{McDonald2022}. This maximum observed luminosity is known as the Humphreys-Davidson (HD) limit \citep{Humphreys1978}. 

It is unclear whether brighter red stars can exist beyond this limit, but earlier work shows that they may \citep{Chen2015,Gilkis2021}. In this case they would be detectable in arcs such as the Spock, in regions with sufficiently high magnification. 
At the redshift of the Spock galaxy, the distance modulus is 42.63 mag. A very luminous star with absolute magnitude $-10$ would have apparent magnitude $\sim 33$--34 (depending on the star temperature and filter and ignoring specific color corrections). If this star is being magnified by a factor $\mu \approx 400$, its apparent brightness would increase by 6.5 mag, making it detectable in standard {\it HST} observations. A lack of detection of such very luminous stars implies that they cannot exist in the volume being amplified by these factors. Since no stars are observed in the Spock arc in several of the observations carried out by {\it HST}, one can use this fact to set an upper limit on the luminosity of the stars in that portion of the arc. During the more frequent periods when the star is not observed, it is logical to assume that the star is being magnified by the most likely value of the magnification. Meanwhile, the star is briefly being observed during relatively short periods when the star is crossing one of the microcaustics in the lens plane. The main objective of this paper is to quantify the probability that a star is being observed during those short periods, but in order to reach that point we first need to spiral down a long path that starts with deriving the mass distribution in the cluster (the macrolens model), follows with the microlens model, then a characterisation of the morphology and stellar properties of the Spock arc, and finally the combination of all the previous ingredients to compute the probability that a star is detected above some established limiting magnitude and for a given filter.  

The paper is organised as follows.
In Section~\ref{Sect_BUFFALO} we briefly discuss the observations carried out during the BUFFALO program as well as the ancillary MUSE data used to derive the spectroscopic redshifts. 
The lensing constraints and data used to derive the lens model are discussed in Section \ref{Sect_constraints}.
Section \ref{Sect_WSLAPplus} gives a brief introduction to the free-form algorithm used to derive the lens model, making no assumptions about the distribution of dark matter (DM). 
In Section \ref{Sect_lensmodel}, we present the lens model derived using only lensed systems with known redshifts (spectroscopic).
Section~\ref{Sect_HighZgal} presents a simple application of the lens model, where we make predictions for the number of observed lensed galaxies around the cluster at $z=6$ and $z=9$. 
  Our attention is focused in Section~\ref{Sect_MicrolensingI} on the macromodel magnification at the position of the Spock arc.
 In Section~\ref{Sect_MicrolensingII}, we study the modifications to the macromodel magnification introduced by microlenses in the lens plane. 
Section~\ref{Sect_Spock} discusses the physical parameters derived for the Spock galaxy. 
 We present the main results from our study in Section~\ref{Sect_Results}, where we combine all of the ingredients from the previous sections and compute the probability of seeing microlensing events in the Spock galaxy. 
Sections~\ref{Sect_discussion} and \ref{Sect_conclusions} respectively discuss these results and present our conclusions.

We adopt a standard flat cosmological model with $\Omega_m=0.3$ and $h=0.7$. At the redshift of the lens ($z=0.396$), and for this cosmological model, one arcsecond corresponds to 5.39 kpc, while at $z_s=1.0054$, one arcsecond corresponds to 8.13 kpc. Unless otherwise noted, magnitudes are given in the AB system \citep{Oke1983}. Common definitions used throughout the paper are the following. We use the term macromodel when we refer to the global lens model derived in Section~\ref{Sect_lensmodel}. The main caustic and main critical curves refer to this model. We use the term micromodel for the perturbations introduced by microlenses in the lens plane, and the term microcaustics for the regions of divergent magnification in the source plane associated with these microlenses. Finally, when referring to the effective temperature of a star, we simply denote this temperature by $T$. 

\begin{figure*} 
   \includegraphics[width=18.2cm]{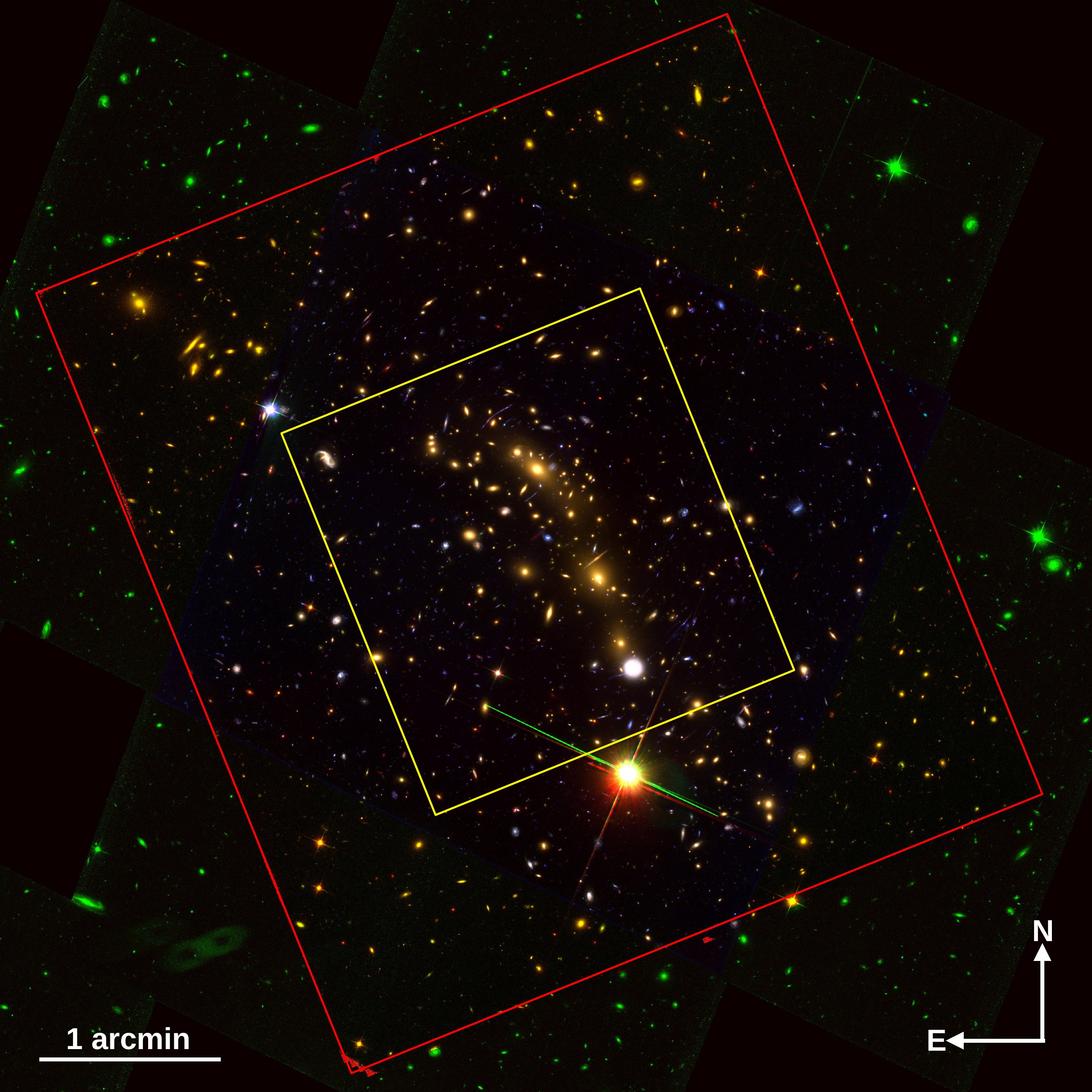}   
      \caption{BUFFALO data around the cluster. The image shown is 6 arcmin on a side. The yellow rectangle marks the region where previous IR data from the HFF program were taken. The red rectangle shows the expanded area covered by BUFFALO in the IR filters. This image corresponds to the combination of filters F435W, F814W, and F160W. Strong-lensing constraints are present only in the central region (yellow square) while weak-lensing constraints are spread over the entire image.  
         }
         \label{Fig_BUFFALO}
\end{figure*}

\section{Observations}\label{Sect_BUFFALO}

In this section, we briefly summarise the high-resolution {\it HST} imaging and the follow-up spectroscopy of MACS0416 that are used to develop the lens model presented in this work. 

\subsection{Hubble Space Telescope imaging}\label{Sect_HST}
The first {\it HST} observations of MACS0416 were performed with WFPC2 (F606W/F814W) in 2007/2008 (GO-11103, PI H. Ebeling). These early high-resolution images of MACS0416 established this cluster as a powerful lens, leading to its selection as one of the six targets in the HFF program \citep{Lotz2017}. 
Thanks to the HFF program, MACS0416 is among the galaxy clusters with the deepest {\it HST} observations obtained to date. The HFF observations have been extended through the \textit{Beyond the Ultra-deep Frontier Fields And Legacy Observations} program \citep[BUFFALO; GO-15117; PIs C. Steinhardt \& M. Jauzac, ][]{Steinhardt2020}. 

The BUFFALO {\it HST} mosaics include the original HFF and archival data, as described by \cite{Steinhardt2020}. The mosaics are produced following the approaches described by \cite{Koekemoer2011}. 
In Figure~\ref{Fig_BUFFALO}, we show the BUFFALO color image of MACS0416. The yellow rectangle highlights the area covered by the IR filters in the original HFF program. The red rectangle shows the larger area covered in the IR bands by the BUFFALO program. Our lens model is constructed over the entire $6\times6$ arcmin$^2$ region shown in this figure. 

\subsection{Spectroscopy}\label{Sect_spectroscopy}
The MUSE observations of MACS0416 used in this work were originally presented as part of the MUSE Lensing Clusters data release \citep{Richard2021} of 12 massive, strong-lensing clusters. Data for MACS0416 were obtained through the MUSE Guaranteed Time Observations (GTO) consortium (PID 094.A-0115, PI J. Richard) and two additional General Observation (GO) proposals (PID 094.A-0525, PI F. Bauer; PID 0100.A-0763, PI E. Vanzella). To maximise the observational depth, all exposures were combined into a master dataset, and for consistency each individual frame was reduced using a common pipeline. Although this procedure largely followed the standard prescription given in the MUSE Data Reduction Pipeline User Manual\footnote{https://www.eso.org/sci/software/pipelines/muse/} based on the method of \citet{Weilbacher2020}, some modifications were made owing to the crowded nature of cluster fields. A full description of the method and its modifications is presented in Section 2.5 of \citet{Richard2021} and Section 2.2 of \citet{Lagattuta2022}. After final combination, the master dataset has a total exposure time of 30\,hr, covering an area of $\sim 2.5$ arcmin$^2$.

Spectroscopic candidates were identified in the field in two ways. The first method (\emph{prior} targets) selected all objects in the combined multiband \emph{HST} imaging using \textsc{Source Extractor} \citep{Bertin1996}, with the size of each detected object slightly broadened [i.e., convolved with the MUSE point-spread function (PSF)] to account for resolution differences. In the second method (\emph{muselet} sources), strong emission-line features were detected in the MUSE data directly, using the \textsc{muselet} routine included in the MUSE Python Data Analysis Framework (MPDAF; \citealt{Piqueras2019}). This allowed for a more complete search of multiple-image systems, identifying objects that were not directly visible in the \emph{HST} data alone. Regardless of the detection method used, a spectrum of each candidate was extracted using an optimally-weighted \citet{Horne1986} algorithm. An initial redshift for each extraction was estimated using the \textsc{MARZ} software \citep{Hinton2016}, though these were later refined by human inspection. A finalised, master list of all redshifts in the MACS0416 MUSE footprint is included in \citet{Richard2021}. 

\section{Strong- and weak-lensing constraints}\label{Sect_constraints}

\begin{figure*} 
   \includegraphics[width=18cm]{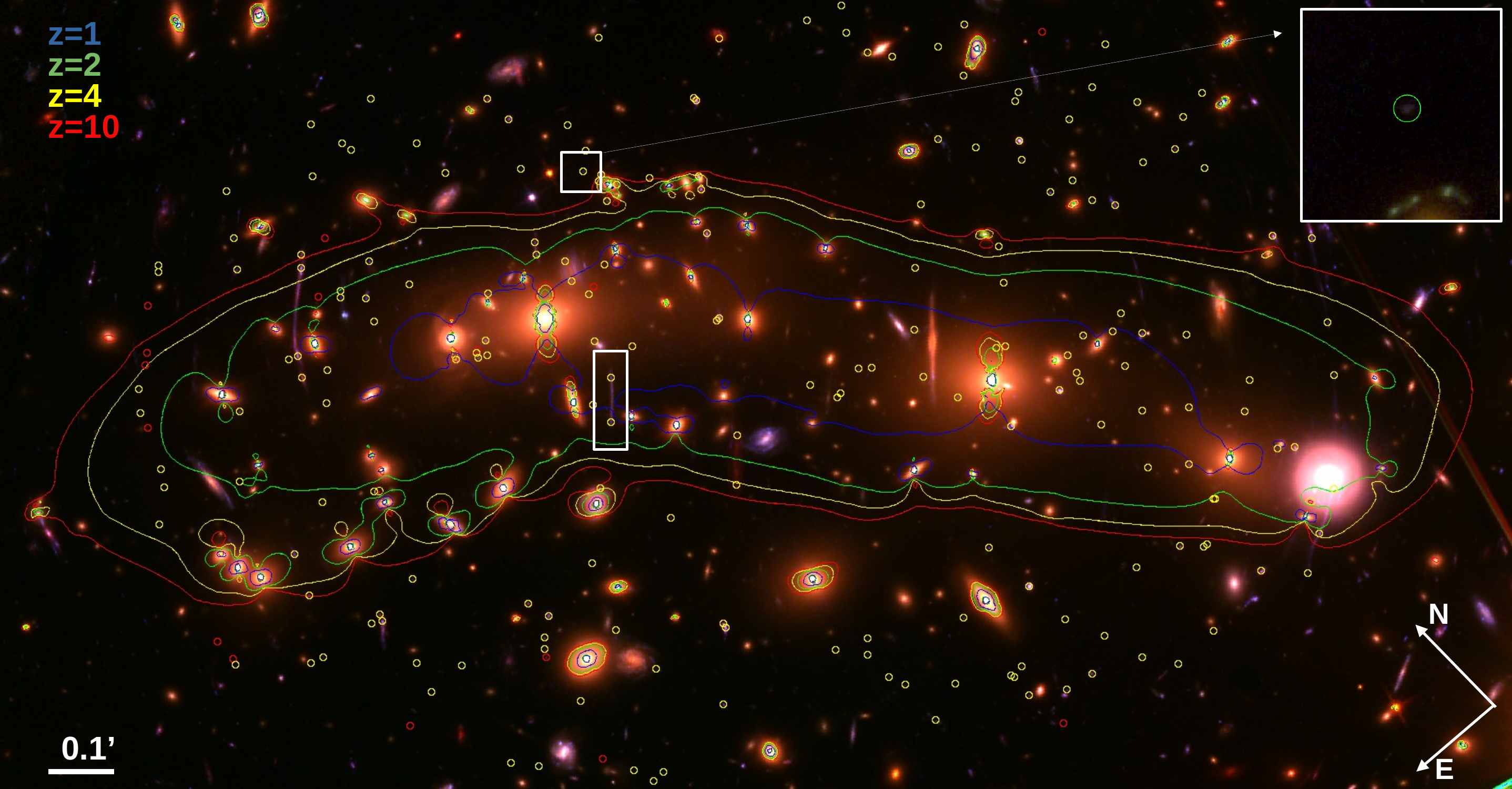}   
      \caption{Central region with arc positions. Yellow circles are the 72 systems identified in BUFFALO data and with spectroscopic redshifts. Red circles are systems not included in the previous sample but listed by \cite{Bergamini2022}. The critical curves for different redshifts are shown ($z=1$, 2, 4, and 10; blue, green, yellow, and red, respectively). This image is a composite made with the seven {\it HST} filters (F435W, F606W, F814W, F105W, F125W, F140W, and F160W). The white rectangle marks the position of the Spock arc. The small white square marks the position of the third counterimage of the Spock arc. A zoomed-in version is shown at top right, where the counterimage is contained within the yellow circle. The radius of this circles is $0.25''$.
         }
         \label{Fig_Arcs}
\end{figure*}

The strong- and weak-lensing catalogs have been compiled within the BUFFALO collaboration following the procedure outlined by \cite{Niemiec2023}. Here we only summarise the main aspects of the process, and refer the reader to that publication for a detailed step-by-step description. The strong-lensing constraints were compiled from the existing literature \citep[][]{Jauzac2014,Johnson2014, Diego2015b, Kawamata2016, Caminha2017, Richard2021} to produce a self-consistent and homogeneous catalog. All multiple images are re-voted within the collaboration and are assigned a grade (gold, silver, or bronze) according to their reliability, with gold systems being the most reliable (spectroscopically confirmed), silver systems being less reliable (they lack spectroscopic confirmation but are good candidates based on consistency in color, morphology, and photometric redshift), and bronze being the least reliable but still potentially correct systems (as for silver systems, bronze systems lack spectroscopic confirmation and are usually unresolved, lacking morphological information). The position of each image is then visually inspected by several members of the collaboration, to ensure that the chosen positions (e.g., the emission knot, brightest pixel, etc.) are consistent within each multiple-image system. In this analysis, we only use the constraints labelled as gold to derive the lens model, which corresponds to 72 multiply lensed galaxies, resulting in 214 images when lensed by the cluster. This can be compared to the recent analysis of \citet{Bergamini2022} which used 60 individual galaxies that are multiply lensed. Two of the  \citet{Bergamini2022} systems in the northeast portion of the cluster are excluded from our analysis, not because they are dubious systems, but simply because our lens models were derived before the results of \citet{Bergamini2022} were published. Adding the two \citet{Bergamini2022} systems would certainly improve our lens model, but not substantially since the region of the lens plane where these two new systems are found is already well constrained by other systems nearby. The locations of our constraints (and the new ones discovered by \citealt{Bergamini2022}) are shown in Figure~\ref{Fig_Arcs}. As seen there, the new systems of \citet{Bergamini2022} (marked as red circles) are not close to the Spock arc, so not adding them in our lens model has negligible impact on our main results. However, future lens models should definitely include these systems, so we add them in our curated list of constraints in the Appendix, and for the convenience of future lens modelling efforts with this cluster. 

The weak-lensing (WL) catalog is produced from the reduced BUFFALO data as follows. The galaxy shapes are measured from their second- and fourth-order normalised multipole moments, using the publicly available \textsc{pyRRG} code \citep{Harvey2021}. All objects (background and foreground galaxies, stars), are first extracted from the mosaic images using two runs of \textsc{Source Extractor} \citep{Bertin1996} with different detection threshold and kernel sizes, in order to best extract both the small and large sources, with the so-called ``hot/cold’’ method. Galaxies and stars are then separated from the image artefacts with a maximum surface brightness ($\mu_{\rm{max}}$) vs. magnitude diagram: the shape of galaxies is what carries the weak-lensing signal, while the stars are used to model the PSF and correct the measured galaxy moments. Cuts are then applied to clean the obtained galaxy-shape catalog, and (i) mask the edges and sources contaminated by the light of massive cluster members or saturated stars, (ii) remove the galaxies that are too small or too faint to have a proper shape measurement, and (iii) remove the sources with ill-measured ellipticities or errors.

The last step is to select galaxies that are located behind the lensing cluster, as foreground galaxies are not weakly lensed and would dilute the measured lensing signal. To perform this separation, we apply colour--colour or colour--magnitude cuts, depending on the number of photometric bands available for each source. Galaxies located within both the BUFFALO ACS and WFC3 fields of view are selected in a $(m_{\rm{814W}} - m_{\rm{160W}}$ vs. $m_{\rm{606W}} - m_{\rm{160W}})$ diagram, while galaxies only covered by ACS observations in a  $(m_{\rm{606W}} - m_{\rm{814W}})$ vs. $m_{\rm{814W}}$ diagram. The cuts are calibrated using the subsample of galaxies that have spectroscopically measured redshifts. The final background galaxy catalog contains 3235 sources. The number density of weak-lensing measurements varies across the field of view. We bin these galaxies in $1\times1$ arcmin$^2$ bins and obtain a mean surface density of 43.6 galaxies per arcmin$^2$ with dispersion $\sigma=18.9$ galaxies per arcmin$^2$. The central region of the cluster contains no WL measurements owing to contamination from the cluster itself, but this region is well constrained by the numerous strong-lensing arcs. We consider a region of $6 \times 6$ arcmin$^2$, hence the number of WL constraints is 72 ($6\times6=36$ bins, each with two constraints $\gamma_1$ and $gamma_2$) .  

\section{WSLAP+}\label{Sect_WSLAPplus}
To optimise the lens model we use the code WSLAP+ \citep{Diego2005,Diego2007,Sendra2014,Diego2016}. A lens model derived using WSLAP+ is considered a hybrid type of model as it combines a free-form decomposition of the lens plane for the smooth large-scale component with a small-scale contribution from the member galaxies. The code combines weak and strong lensing in a natural way, with changes made in the large-scale and small-scale components impacting equally the strong- and weak-lensing observables. Details can be found in the references above. Here we give a brief description of the method.

We start with the classic definition of the lens equation
\begin{equation} 
\beta = \theta - \alpha(\theta,\Sigma) \, , 
\label{eq_lens} 
\end{equation} 
where $\theta$ is the observed position of the source, $\alpha$ is the deflection angle, $\Sigma(\theta)$ is the unknown surface mass density of the cluster at the position $\theta$, and $\beta$ is the unknown position of the background source. 
The optimisation of the WSLAP+ solution takes advantage of the fact that the lens equation can be expressed as a linear function of the surface mass density, $\Sigma$. WSLAP+ parameterises $\Sigma$ as a linear superposition of functions, which translates into $\alpha(\theta,\Sigma)$ also being linear in $\Sigma$. 

In WSLAP+, $\Sigma$ is described by the combination of two components: 
(i) a smooth component (usually parameterised as superposition of Gaussians) corresponding to the free-form part of the model, or large-scale cluster potential, and 
(ii) a compact component that accounts for the mass associated with individual galaxies in the cluster.
For the smooth component, we use Gaussian functions defined over a grid of points. A Gaussian function is simple, and enables fast computation of the deflection field, but also provides a good compromise between the desired compactness and smoothness  of the basis function. 
The grid configuration can be defined as regular (all grid points have the same size) or irregular (grid points near the centre are in general smaller). Adopting a regular grid is similar to a flat prior in the mass distribution, while an irregular grid can be interpreted as a model with a prior on the mass distribution with higher mass density assigned to smaller cells. For this work, we use a dynamical (irregular) grid with increased resolution in the central region, where strong-lensing constraints are present. In particular, we consider a grid with 745 points. The distribution of grid points is obtained from a Monte Carlo realisation of the mass density previously obtained  with a regular grid. The final distribution of points is more concentrated around the two brightest cluster galaxies (BCGs), with the density of grid points being smallest in the outskirts of the $6 \times 6$ arcmin$^2$ region. 

For the compact component, we directly adopt the light distribution in {\it HST}'s F160W band around the brightest member elliptical galaxies in the cluster, and spectroscopically confirmed (by MUSE) members. 
For each galaxy, we assign a mass proportional to its surface brightness. This mass is the only free parameter, and it is later   readjusted as part of the optimisation process. The number of parameters connected with the compact component depends on the number of adopted layers. Each layer contains a number of member galaxies. The minimum number of layers is 1, corresponding to the case where all galaxies are placed in the same layer --- that is, they are all assumed to have the same mass-to-light ratio. In this case, the single layer is proportional to the light distribution of all member galaxies, and is assigned a fiducial mass accounting for the total mass of member galaxies. For each layer there is one extra parameter which accounts for the renormalisation constant multiplying the map of the mass distribution. This renormalisation parameter is later optimised by WSLAP+. For the particular case of MACS0416, we use four layers. The first two layers contain the two main BCGs, the third layer contains red-sequence member galaxies, and the fourth layer contains spectroscopically confirmed member galaxies which are not in the red sequence and hence may have a different mass-to-light ratio. 

As shown by \cite{Diego2005,Diego2007}, the strong- and weak-lensing problem can be expressed as a system of linear equations that can be represented in a compact form, 
\begin{equation}
\vectTheta = \matrGamma \vectX, 
\label{eq_lens_system} 
\end{equation} 
where the measured strong- and weak-lensing observables are arranged in $\vectTheta$ of dimension $N_{\Theta }=2N_{\rm sl} + 2N_{\rm wl}$ (where $N_{\rm sl}$ and $N_{\rm wl}$ are the strong- and weak-lensing observables, each contributing with two constraints). The unknown surface mass density and source positions are in the array $\vectX$ of dimension 
\begin{equation}
N_{\rm X}=N_{\rm c} + N_{\rm l} + 2N_{\rm s}\,. 
\label{eq_Nx}
\end{equation}
The  array $\vectX$ also contains the $N_{\rm l}=4$ unknown renormalisation factors associated with the four layers. $N_{\rm c}$ is the number of grid points (or cells) that we use to divide the field of view,  and $N_s$ is the number of background sources being strongly lensed (each source contributes with two variables, $\beta_x$ and $\beta_y$).  
Finally, the matrix $\matrGamma$ is known (for a given grid configuration and fiducial galaxy deflection field) and has dimension $N_{\Theta }\times N_{\rm X}$.


The solution, $X$, of the system of Equations \ref{eq_lens_system} is found after minimising a quadratic function of $X$ \citep[derived from the system of Equations \ref{eq_lens_system} as described by ][]{Diego2005}.
A detailed discussion of the algorithm used to solve the system of linear equations is given by \cite{Diego2005}. For a discussion of its convergence and performance (based on simulated data), we refer the reader to \cite{Sendra2014}.

\section{The derived lens model}\label{Sect_lensmodel}

We derive our lens model combining weak- and strong-lensing constraints. The resulting model shows the characteristic bimodal distribution found in earlier work, with two peaks of mass centred in each of the BCG galaxies.  
The critical curves for $z=1$, 2, 4, and 10 are shown in Figure~\ref{Fig_Arcs}. The critical curves are very elongated and intersect multiple arcs at their respective redshifts. These intersection points offer unique opportunities to zoom in the highly magnified background galaxies, as discussed in more detail in Section 6. 

\begin{figure} 
   \includegraphics[width=9cm]{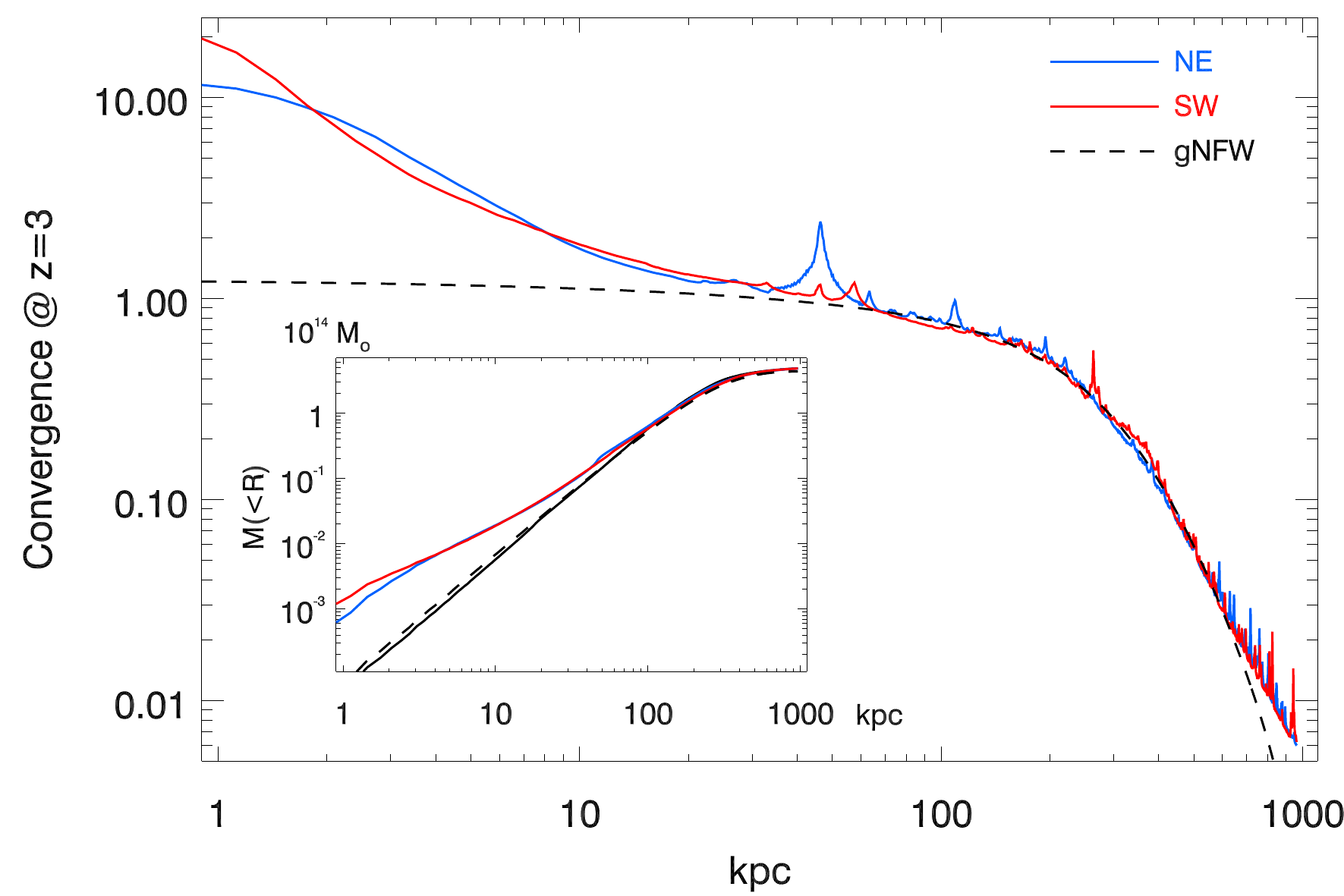} 
      \caption{Mass profile of the lens model. The blue  and red solid curves show the profiles when adopting the NE and SW BCGs as central points, respectively. The mass profiles are given in dimensionless $\kappa$ units assuming a source at $z=3$. The dashed line is a gNFW profile with parameters $\gamma=0.5$, $\alpha=2.5$, $\beta=3$, and scale radius $r_s= 80$\,kpc. The inset shows the integrated mass as a function of radius in units of $10^{14}\, \Msun$, and for the same three profiles shown in the larger plot. The black dashed line is the integrated profile when the centre is chosen as the middle point between the two BCG galaxies (RA = $4^{\rm hr} 16^{\rm m} 08.428^{\rm s}$, Dec. = $-24^\circ 04' 21.0''$). 
         }
         \label{Fig_Profile}
\end{figure}

Regarding the mass profile, we find that the two main clumps of mass are very similar to each other in terms of density profiles. These are shown in Figure~\ref{Fig_Profile}. We compute one profile centred in the northeast (NE) BCG galaxy and a second profile centred in the southwest (SW) BCG galaxy. The profiles represent the projected mass along the line of sight in a cylindrical aperture. The most remarkable difference between the two profiles is that the SW BCG is more massive within the central kpc, but beyond a few kpc both profiles are very similar. We fit a generic Navarro-Frenk-White profile (gNFW), finding a reasonable fit for a model with parameters $\gamma=0.5$, $\alpha=2.5$, $\beta=3$, and scale radius $r_s= 80$\,kpc. The fit is done centering on each halo so it does not account for superposition effects of one halo over the other. This profile fails at reproducing the inner $\sim 20$\,kpc region, but this is where baryons are expected to cool more efficiently than dark matter and form a  baryonic cusp. At larger radii, our lens model benefits from the addition of weak-lensing data that extends the range of constraints up to the boundaries of the field of view shown in Figure\ref{Fig_BUFFALO}. 

In terms of integrated mass as a function of radius, we show the cumulative mass profile in the inset of the same figure. The blue and red curves are again for the NE and SW clumps, while the dashed line is for the gNFW profile. In this plot we add a fourth curve (black solid line) that corresponds to a third profile centred in the middle point between the two BCG galaxies (the separation between these galaxies is $\sim 40''$). We find that the total mass contained within 1\,Mpc for the black continuous line is $4.9\times10^{14}\, \Msun$. 
\cite{Bergamini2022} derive a parametric model of this cluster and show how the integrated mass for each halo is also very similar to each other. Quantitatively, from their Figure 7, they demonstrate how the integrated cylindrical mass around each halo, measured at a radius of 93\,kpc, is  $\sim 4.8\times10^{13}\,\Msun$. For the same radius, we find masses $5.6\times10^{13}\, \Msun$ and $5.2\times10^{13}\, \Msun$ for the NE and SW halos (respectively), which is 5\%--10\% higher. At larger radii, there is no quantitative measure provided by \cite{Bergamini2022}, but from the same plot they find a mass for each halo of $\sim 3.4\times10^{14}\, \Msun$ at $R=400$\,kpc. At this distance we find  $3.6\times10^{14}\, \Msun$ for each, the NE and SW halos.

\section{Expected number of multiply lensed galaxies at high redshift in JWST images}\label{Sect_HighZgal}
\begin{figure} 
   \includegraphics[width=9cm]{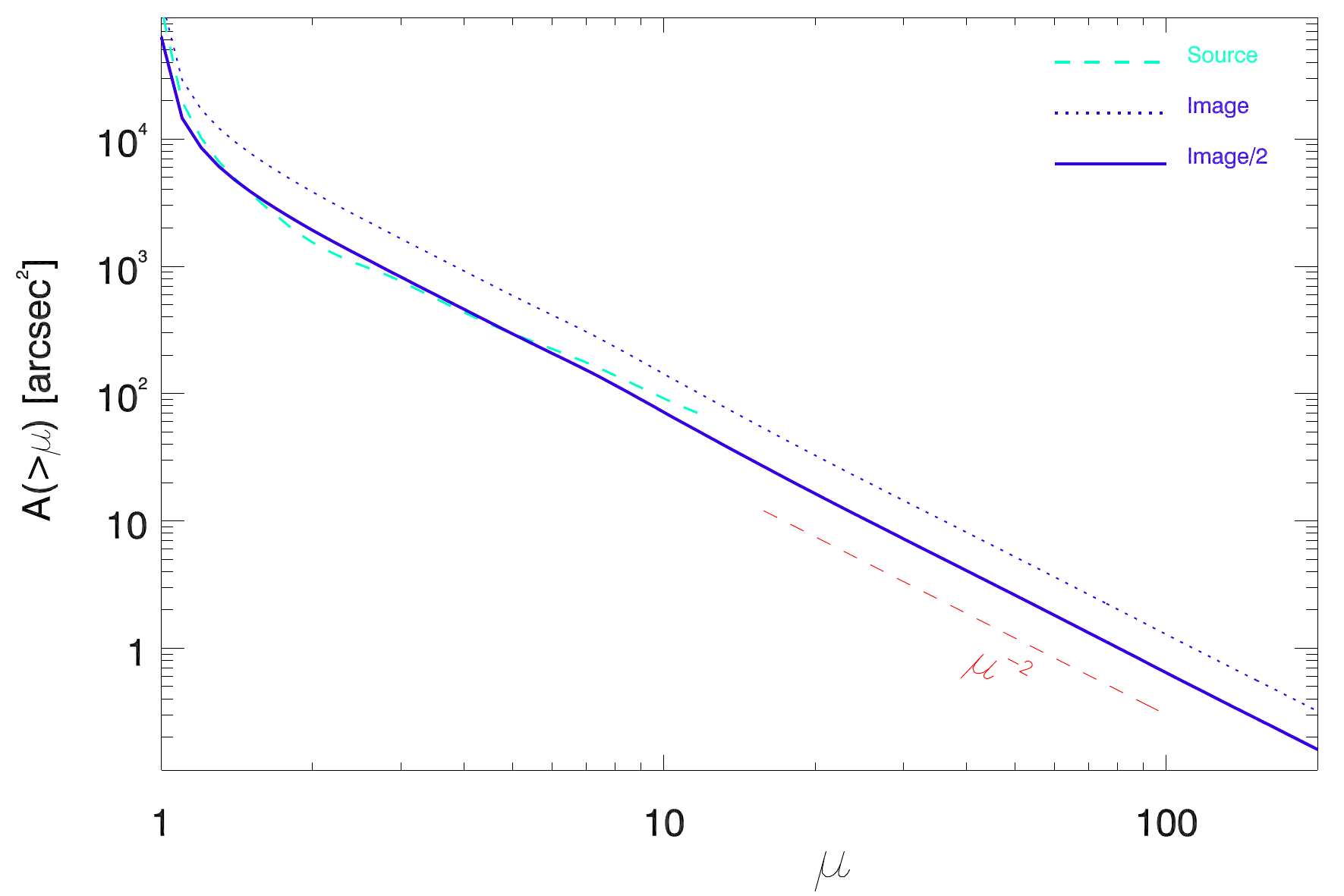} 
      \caption{Area above magnification $\mu$ at $z=9$. The dashed line is computed in the source plane from the caustic map. The dotted line is the area computed in the image plane from the magnification map corrected by the factor $\mu$. The solid line is similar to the dotted line but divided by a factor of 2 (in order to account for the contribution from the double image at large magnification factors). At large magnification factors, the solid curve falls as the expected $\mu^{-2}$ power law. 
         }
         \label{Fig_Mu}
\end{figure}

In this section, we estimate the expected number of lensed galaxies observed at $z=6$ and $z=9$, based on the magnification prediction from the lens model.  For sources at $z=6$ the corresponding curve $A(>\mu)$ is almost identical to that shown in Figure~\ref{Fig_Mu}, with the amplitude of this curve being just $\sim 14\%$ smaller than the curve for $z=9$ \citep[based on the scaling with $z$ of the lensing strength factor, Eq. 2.4 in][]{SeitzSchneider1997}.
Here we consider only the case of multiple images, which takes place when the total magnification is above $\mu \approx 3$. Outside the caustic region the total magnification is below 2.5, so this assumption is well justified.

At a given redshift, the number of lensed galaxies can be expressed in the form
\begin{equation}
 \frac{{\rm d}N}{{\rm d}\mu \, {\rm d}M_{\rm UV}} = \Phi(z,\tilde{M}_{\rm UV})*dV(z)N_{ci}\frac{dA}{d\mu}, 
 \end{equation}
 where $M_{\rm UV}$ is the absolute UV luminosity of the galaxy and $\tilde{M}_{\rm UV}= M_{\rm UV} - 2.5\,log_{10}(\mu)$ is the magnitude before magnification. $\Phi(z,M_{\rm UV})$ is the luminosity function at redshift $z$ (expressed in the usual units of number per magnitude per Mpc$^3$), and ${\rm d}A/{\rm d}\mu$ is the area in the source plane (expressed in arcseconds squared) with total magnification $(\mu-0.5)<\mu<(\mu+0.5)$. Since the area $dA/d\mu$ is given in arcsec$^2$, the volume  $dV(z)$ corresponds to the volume within 1  arcsec$^2$ and $(z-0.5)<z<(z+0.5)$. Finally, $N_{ci}$ is the number of detectable counterimages produced by each lensed galaxy. We make the simple approximation that the total magnification $\mu$ of the galaxy is distributed into three counterimages. This is a general statement which is valid for most lensed galaxies. When the galaxy is in the central valley of the cluster caustic region, the three counterimages have similar magnification. The central valley for MACS0416 has typical magnification $4<\mu<12$. Hence, we assume that galaxies falling in this central-valley region form three counterimages, each having magnification $\mu/3$. To differentiate from the valley region, we define the region with magnification $\mu>12$ as the peak region. For typical clusters, the peak region is just a few kpc away from the caustics. For galaxies in the peak region, there are still three counterimages, but for galaxies near fold caustics (the vast majority), one of the counterimages typically has smaller magnification factors than counterimages produced in the valley region, while the other two counterimages carry most of the magnification. Hence, we do the additional approximation that galaxies located in the valley region form three detectable images ($N_{ci}=3$), with moderate magnification factors, while galaxies in the peak region form two detectable images ($N_{ci}=2$) with large magnification factors. 

 For the luminosity function of background galaxies, $\Phi(z,M_{\rm UV})$, we follow \cite{Bowler2015} for $z=6$ and \cite{Harikane2022,Pablo2023} for $z=9$. 
 Using these luminosity functions, we consider three different types of observations: (i) shallow, Where observations can detect galaxies brighter than $M_{\rm UV}=-17$\,mag; (ii) medium, Where galaxies with $M_{\rm UV}=-16$\,mag can be detected; and (iii) deep, reaching down to $M_{\rm UV}=-15$\,mag. These absolute magnitudes correspond to those before correcting for amplification. That is, if a galaxy is inferred to have $M_{\rm UV}=-15$\,mag, but is magnified by a factor $\mu=10$, the magnification-corrected magnitude would be $M_{\rm UV}=-12.5$\,mag.  The distance modulus for a source  at $z=6$ and $z=9$ is 44.63 and 44.87\,mag, respectively. Hence, a galaxy with $M_{\rm UV}=-15$\,mag at $z=9$ would have an apparent magnitude (in the IR) of 29.87, within reach of deep observations made by {\it JWST} (ignoring colour corrections and extinction).

\begin{figure} 
   \includegraphics[width=9cm]{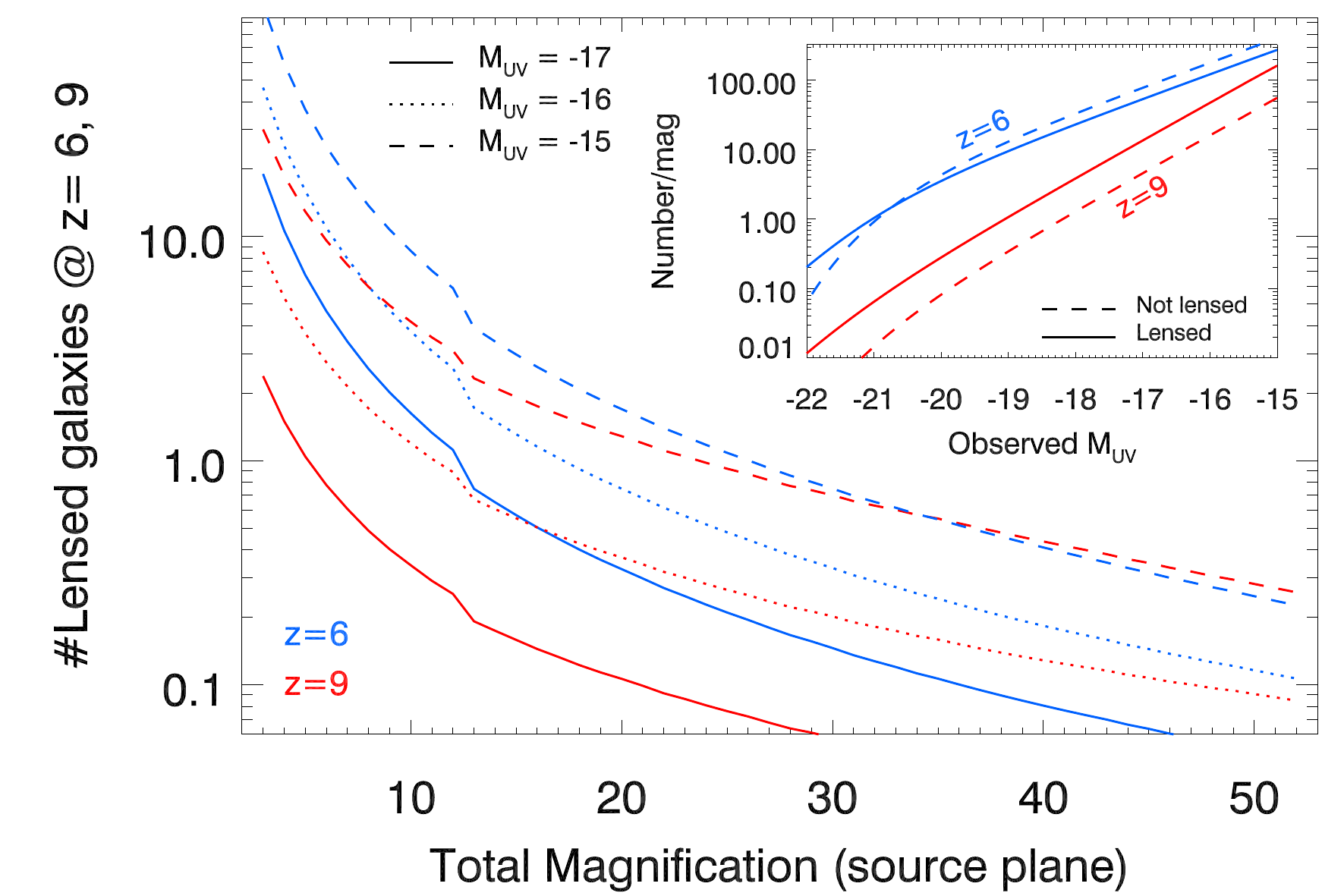} 
      \caption{Expected number of lensed galaxies in MACS0416. The larger plot shows the number of lensed galaxies behind MACS0416 at $z=6$ (blue curves) and $z=9$ (red curves), as a function of magnification (larger plot) or observed absolute magnitude (smaller plot). We assumed a standard luminosity function at each redshift. The different lines are for different limiting magnitudes: $-17$ (solid lines), $-16$ (dotted lines), or $-15$ (dashed lines). In the smaller plot the solid lines are the number of lensed galaxies for each observed magnitude, while the dashed lines are the number of galaxies one would have observed in the same magnitude bins but without lensing magnification.  
         }
         \label{Fig_NLensedGal}
\end{figure}


The results are summarised in Figure~\ref{Fig_NLensedGal}. In the larger plot, we show the number of lensed galaxies as a function of their total magnification --- that is, the magnification experienced by the galaxy in the source plane. 
As mentioned earlier, in the image plane, each lensed galaxy usually produces between two and three detectable counterimages.\footnote{More counterimages may appear owing to substructure or when the galaxy is also producing radial images, but this situation is not considered here.} 

In our case, the transition between two and three detectable counterimages takes place when the total magnification is $\mu=12$. This transition at $\mu=12$ can be easily appreciated in Figure~\ref{Fig_NLensedGal}, which
shows the results obtained for the two redshifts ($z=6$ and 9) and the three different depths (from bottom to top: $M_{\rm UV}=-17$, $-16$, and $-15$\,mag).
As expected, most lensed galaxies have relatively small magnification factors simply because smaller magnification factors correspond to larger areas in the source plane. 
For the deepest observations ($M_{\rm UV}=-15$\,mag), we expect $\sim 1$ strongly lensed galaxy with total magnification $\mu\approx30$. This is a galaxy near a caustic or even with parts of it crossing a caustic. Large magnification factors are more likely for high-redshift galaxies owing to their smaller intrinsic size. Recent results from {\it JWST} reveal very small sizes for faint galaxies at  $z \approx 9$ with estimated sizes a few tens of parsecs \citep{Hayley2023}. These galaxies appear unresolved even in {\it JWST} images. With tangential magnification factors of order 10, these galaxies can be resolved, so we expect that one such galaxy may be detectable in deep observations of MACS0416 with {\it JWST}. Surprisingly, from the same figure, this is the same number of galaxies we expect to observe with the same magnification factor, but at $z=6$. This is due to the fact that the adopted luminosity function is steeper at $z=9$ than at $z=6$. This is better appreciated in the inset of Figure~\ref{Fig_NLensedGal}, where we plot the number of detected galaxies (now corrected for multiplicity) as a function of the survey depth. Our model predicts that $\sim 100$ galaxies should be detectable in deep {\it JWST} observations both at $z=6$ and $z=9$. 

In the inset of Figure~\ref{Fig_NLensedGal}, we show as a dashed line the expected number of galaxies in a blank field (that is, with magnification $\mu=1$) of similar area to the region where counterimages are expected to form in MACS0416. For galaxies at $z=6$ we predict the classic depletion effect, where one is less likely to find faint galaxies in lensing fields than in blank fields. In this case, lensing increases the number of detections only at the bright end. However, for $z=9$ galaxies, one is expected to be more successful (by a factor $\sim 3$) in the cluster field than in the parallel field. A similar conclusion was reached by \cite{Mahler2019} for galaxies at $z=10$. These predictions can be tested with ongoing observations of MACS0416 that already cover the cluster region and several parallel fields. If the number density of galaxies found in the parallel fields is greater than in the cluster field, this would be a clear indication that the slope of the luminosity function must be shallower than 2 at the faint end (that is, for galaxies fainter than $M_{\rm UV}=-15$\,mag).

\section{Microlensing events in the Spock galaxy}\label{Sect_MicrolensingI}
In this section, we focus on one of the lensed galaxies behind MACS0416. Nicknamed Spock by \cite{Rodney2018}, it is at relatively low redshift ($z=1.0054$), but crosses one of the fold caustics of the cluster and hence is being significantly stretched and amplified. 
Spock harbors several alleged bright stars\footnote{See \cite{Rodney2018} for possible alternative interpretations}, some of which are close enough to the cluster caustic and can be detected individually when undergoing microlensing events. Two of these microlensing events were reported by \cite{Rodney2018} and two additional events by \cite{Kelly2023}. With {\it JWST} observations of this cluster becoming public soon, revealing more details about this interesting arc, here we pay special attention to this galaxy and the bright stars in its interior.

\begin{figure} 
      \includegraphics[width=9cm]{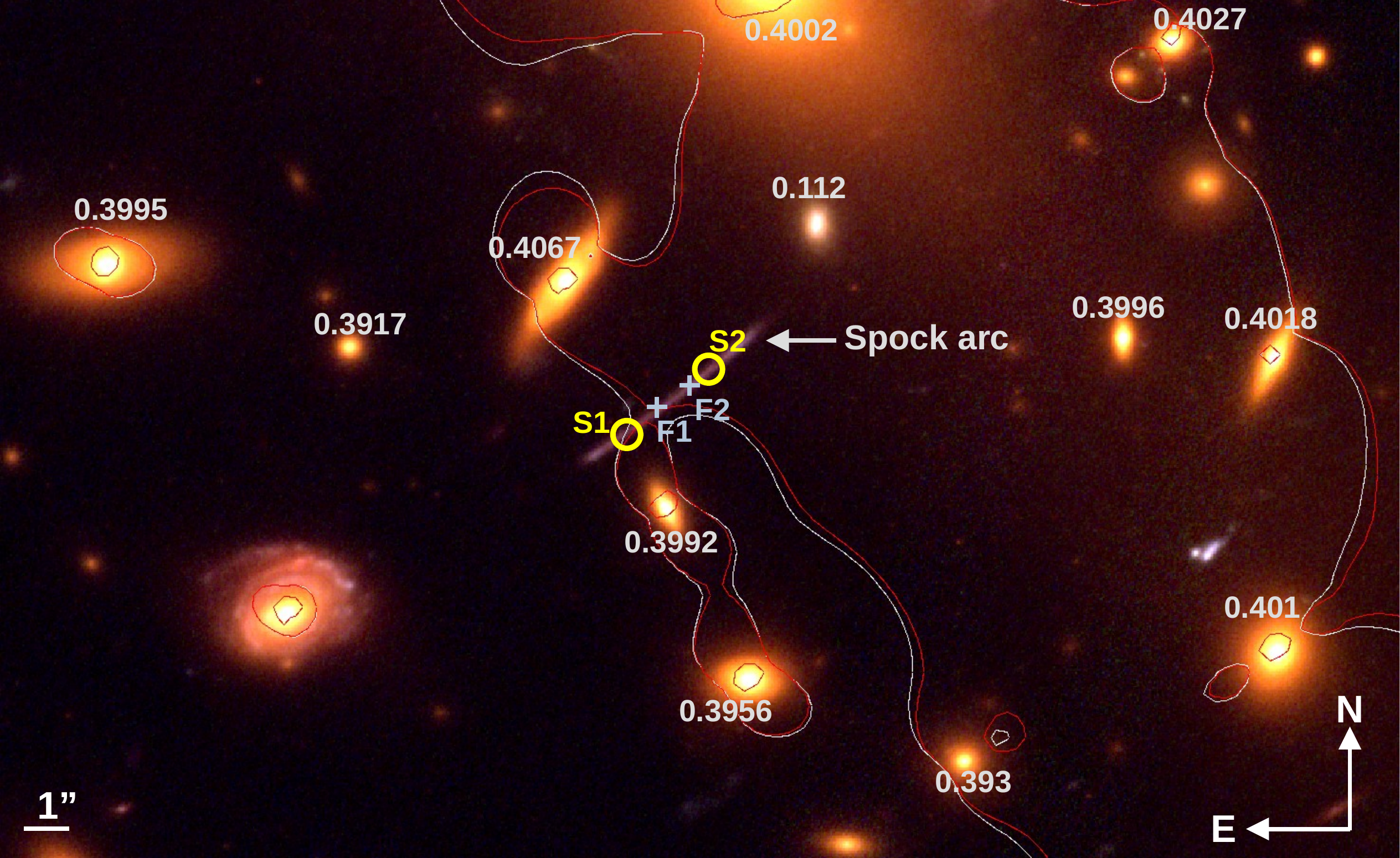}
      \caption{The Spock arc. The radial arc at $z=1.0054$ shows the location of the two Spock transients (S1 and S2 in yellow circles) reported by \cite{Rodney2018}. Two additional transients are also marked with light-blue crosses (F1 and F2). These were discovered in the same arc as part of the Flashlights program \citep{Kelly2023}. Redshifts of nearby galaxies are indicated in light grey. The white line is the critical curve at $z=1.0054$ assuming the cluster is at $z=0.396$. The red curve assumes the cluster is at a slightly larger redshift of  $0.4$ and consistent with these nearby galaxies.
         }
         \label{Fig_SpockCritCurves}
\end{figure}

\begin{figure*} 
   \includegraphics[width=9cm]{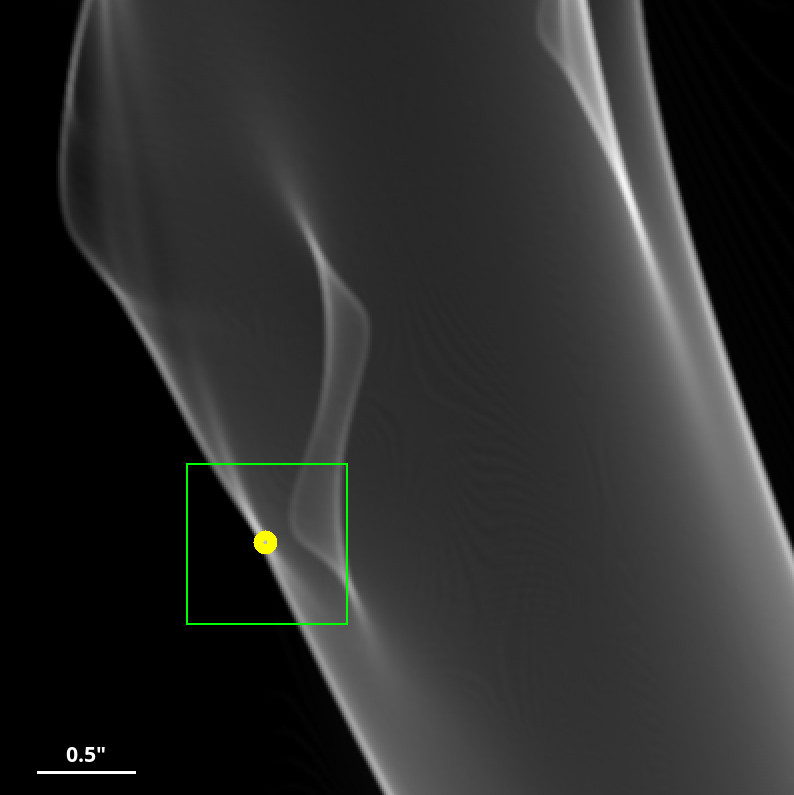} 
   \includegraphics[width=9cm]{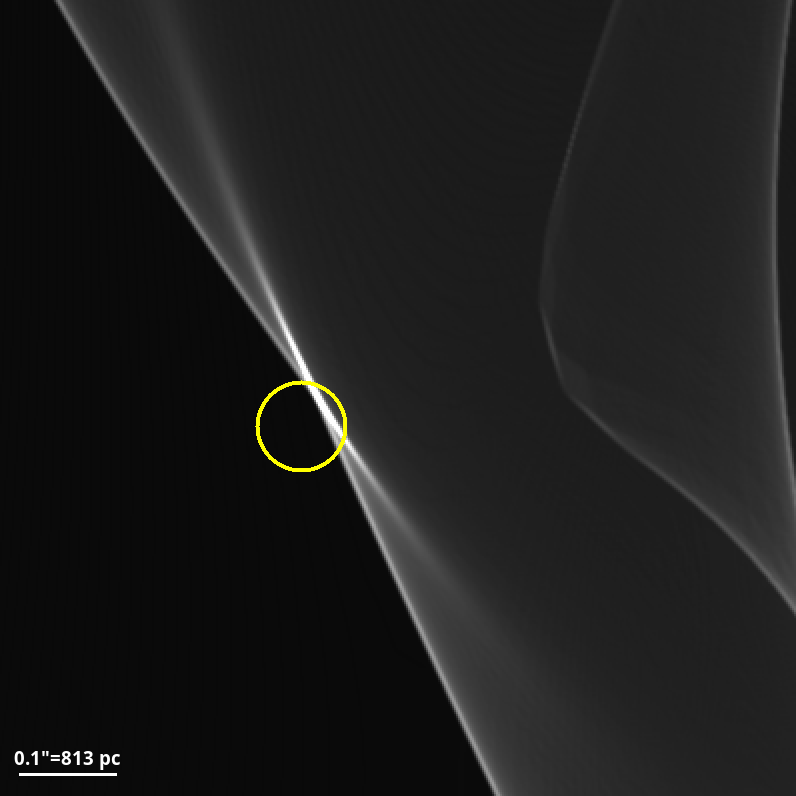} 
      \caption{Portion of the caustic region with the magnification shown as grey scale. {\it Left:} small region of $4'' \times 4''$ showing the caustics for a source at $z=1.0054$. The yellow circle marks the predicted position of the Spock galaxy in relation to the caustics. The green square marks the region that is enlarged in the right panel. {\it Right:} Zoom-in of a small portion in the caustics obtained after interpolating the deflection field to a resolution of 2 milliarcseconds per pixel and applying the inverse ray shooting method. The yellow circle represents the approximate circular shape of the Spock galaxy and its position in relation to the caustic. The length of the caustic that intersects the galaxy is estimated to be $\sim 530$\,pc. At a distance $d$\,pc from the caustic, the magnification in the interior region of the caustic drops as $\mu \approx 270/\sqrt{d\,({\rm pc})}$.}
         \label{Fig_SpockCaustics}
\end{figure*}

The elongated arc from the Spock galaxy is shown in the centre of Fig.~\ref{Fig_SpockCritCurves}. The lensed image forms a long arc perpendicular to the main axis and the critical curve of the cluster. A third image (with a much smaller magnification $\mu\approx 3.5$ and shown in the upper-right corner of Fig.~\ref{Fig_Arcs}) is found in the northern part of the cluster (and outside the field of view shown in Fig.~\ref{Fig_SpockCritCurves}). The elongated arc is in fact a merging image showing only a small portion of the Spock galaxy. Several lens models are presented by \cite{Rodney2018} showing how this merged image is the superposition of at least two counterimages, but possibly more than two. 

The existence of the transients S1 and S2 (labelled in the image and originally reported by \cite{Rodney2018}) supports the existence of multiple caustic crossings. This is further supported by the discovery of two new transients by \cite{Kelly2023} as part of the Flashlights program (and marked  F1 and F2 in the figure) near the position of the two original transients. \citet{Kelly2023} suggest that S1--F1 form an image pair of the same star, while S2--F2 form a different pair of another star. This would imply that one critical curve must pass through S1--F1 and a second one through S2--F2, making the arc a superposition of at least three counterimages. The critical curve at $z=1.0054$ from our new lens model is shown in white in Figure~\ref{Fig_SpockCritCurves}. As expected, this critical curve intersects the arc, but it does it only once, and very close to the point between S1 and F1. The points S2 and F2 do not have a critical curve passing through them, but they are $< 1''$ away from one. Small changes in the mass of the nearby member galaxies, an invisible millilens (for instance, a dwarf galaxy in the cluster), or the large-scale mass distribution, can modify this portion of the critical curve by a small amount, making a second crossing through S2--F2 possible. We show in red an alternative lens model where the redshift of the cluster is increased slightly from the originally adopted $z=0.396$ to $z=0.4$. The increase in redshift in this portion of the lens plane is motivated by the existence of member galaxies nearby at redshifts close to (or even larger than) $z=0.4$, including the BCG to the north which is also slightly above $z=0.4$. Interestingly, this alternative model places the critical curve closer to S2--F2, making the hypothesis of a second caustic crossing more plausible.  

The magnification seen by the Spock galaxy is better shown in the source plane. From the lens model we derive a map of the magnification in the source plane around the predicted position of the Spock galaxy. At the redshift of this galaxy, $1'' = 8.13$\,kpc. In order to increase the spatial resolution, we interpolate the smooth deflection field before applying a standard inverse ray shooting algorithm to compute the caustics. The interpolated region is sufficiently large to guarantee that the inverse ray shooting method is not missing pixels in the image plane that land near the predicted position of the Spock galaxy. The resulting caustics are shown in Figure \ref{Fig_SpockCaustics}. From this result we find that the Spock galaxy crosses a caustic, as expected from the fact that two of its counterimages form a merging pair of images. When zooming in around the Spock galaxy, we find that several caustics converge in this region. This explains the unusually large elongation of the main arc, despite the small intrinsic size of the galaxy. In the caustic map it is also evident that small changes in the lens configuration can result in multiple caustic crossings for a source placed in the right position.  The magnification at the position of Spock scales as $\mu \approx 3/\sqrt{d}$, where $d$ is the distance to the caustic expressed in arcseconds, or $\mu \approx 270/\sqrt{d}$ when $d$ is expressed in parsecs. Using this scaling, we can estimate the magnification at different distances from the caustic.

\section{The role of microlenses}\label{Sect_MicrolensingII}
The macromodel magnification represents only the expected value of the magnification in a given region of the source plane. In realty, the ubiquitous presence of microlenses in the image plane creates a network of microcaustics in the source plane. A new magnification pattern emerges from reshuffling the macromodel magnification, and transforming its distribution from a delta function (the macromodel value) to a broad probability distribution.   
Microcaustics from microlenses in the cluster can temporarily intersect a star in the  background galaxy, providing an extra boost in magnification and making these stars detectable. In order to test the likelihood of detecting the star through a microlensing event, we have to calculate the probability of intersecting microcaustics. 

The simulation with microlenses involves two main ingredients: (i) the macromodel, which is described by the tangential and radial magnifications (or similarly by $\kappa$ and $\gamma$), and (ii) the micromodel, which is basically given by the surface mass density of microlenses (we assume a standard Salpeter IMF for the distribution of masses with a lower cutoff in mass at $0.02\, \Msun$). 
For the macromodel we use the values of $\kappa=0.691$ and $\gamma=0.303$ obtained at the position S1 from the macromodel.  
The exact values of $\kappa$ and $\gamma$ are not relevant as long as they are representative of the values expected around this position. This results in magnification $\mu=\mu_t\mu_r=166.7\times1.64=273.4$, which is in the range of magnifications expected for S1. To simulate different values of the magnification we simply change $\kappa$ very slightly until the desired magnification is reached. 

For the microlens model, we assume that the contribution from the intracluster light (ICL) to the total mass at the distance of  S1 from the main BCG ($\sim 50$\,kpc) is $\sim 2\%$, or $\kappa_*=0.01\,\kappa=0.01382$. 
This number is in agreement with recent results \citep{Montes2022b,Diego2023}. However, given the bimodal morphology of MACS0416, and the possible recent merger activity, the ICL could be even higher, so one should consider the 2\% contribution as a conservative limit. This possibility is reinforced by the fact that the ICL is still clearly visible in {\it HST} images at the position of S1.  Since the critical surface mass density for the redshifts of the cluster and the Spock galaxy is $\Sigma_{\rm crit}=2818\, M_{\odot}$\,pc$^{-2}$, the surface mass density of microlenses corresponds to $\Sigma_* = 19.45\,\Msun\,{\rm pc}^{-2}$. This estimate is in good agreement with the upper limit for the ICL derived by \cite{Rodney2018}. They find $3.2\,\Msun\,{\rm pc}^{-2} < \Sigma_* < 19.4\,\Msun\, {\rm pc}^{-2}$ for the position of the Spock elongated arc. 
In this work we do not consider exotic forms of dark matter that could also act as microlenses, such as primordial black holes (PBHs). Earlier work based on microlensing of distant stars has constrained the abundance of PBHs to a small fraction ($<10$\%) of the total projected mass \citep{Oguri2018}.

\begin{figure*} 
   \includegraphics[width=9cm]{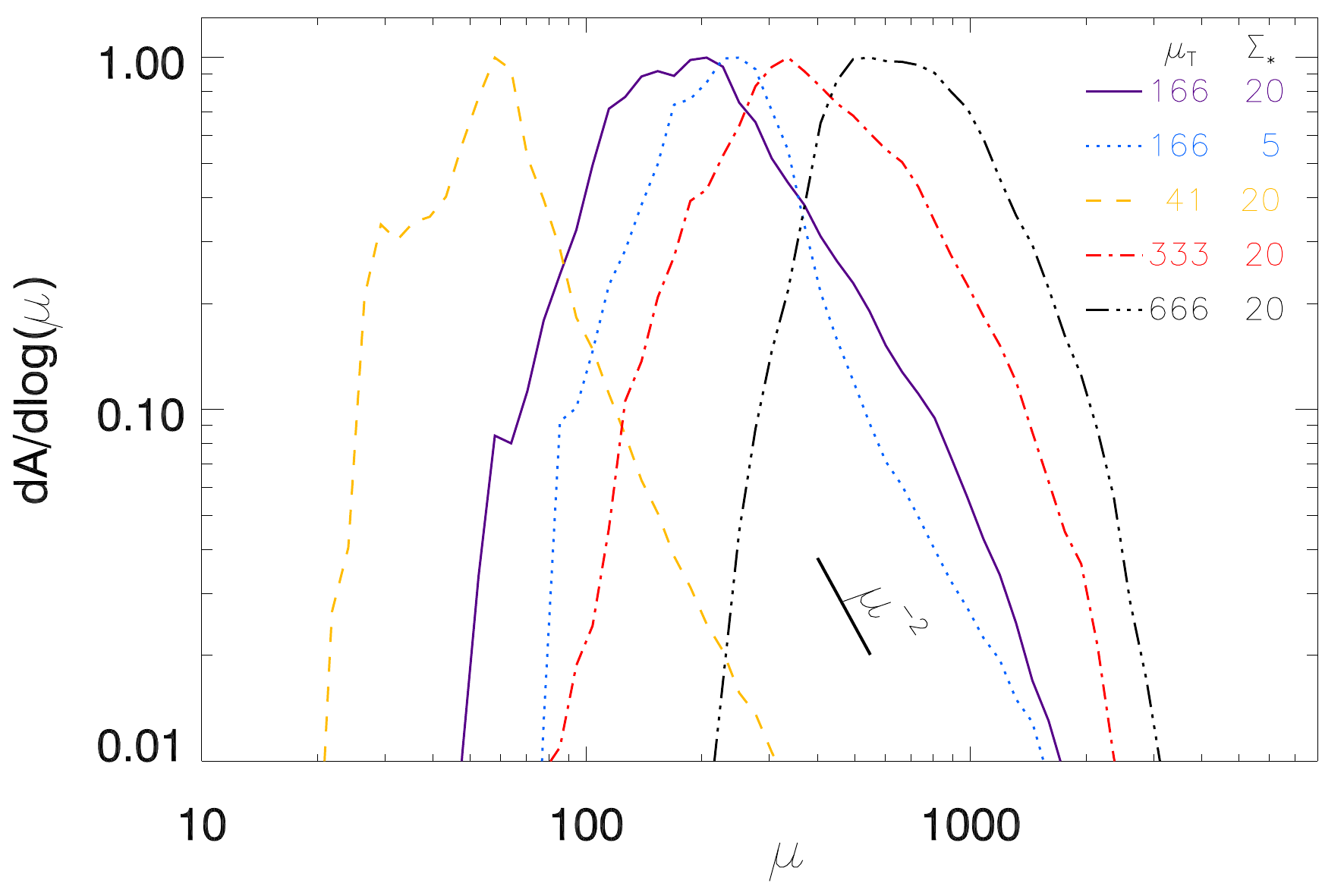} 
   \includegraphics[width=9cm]{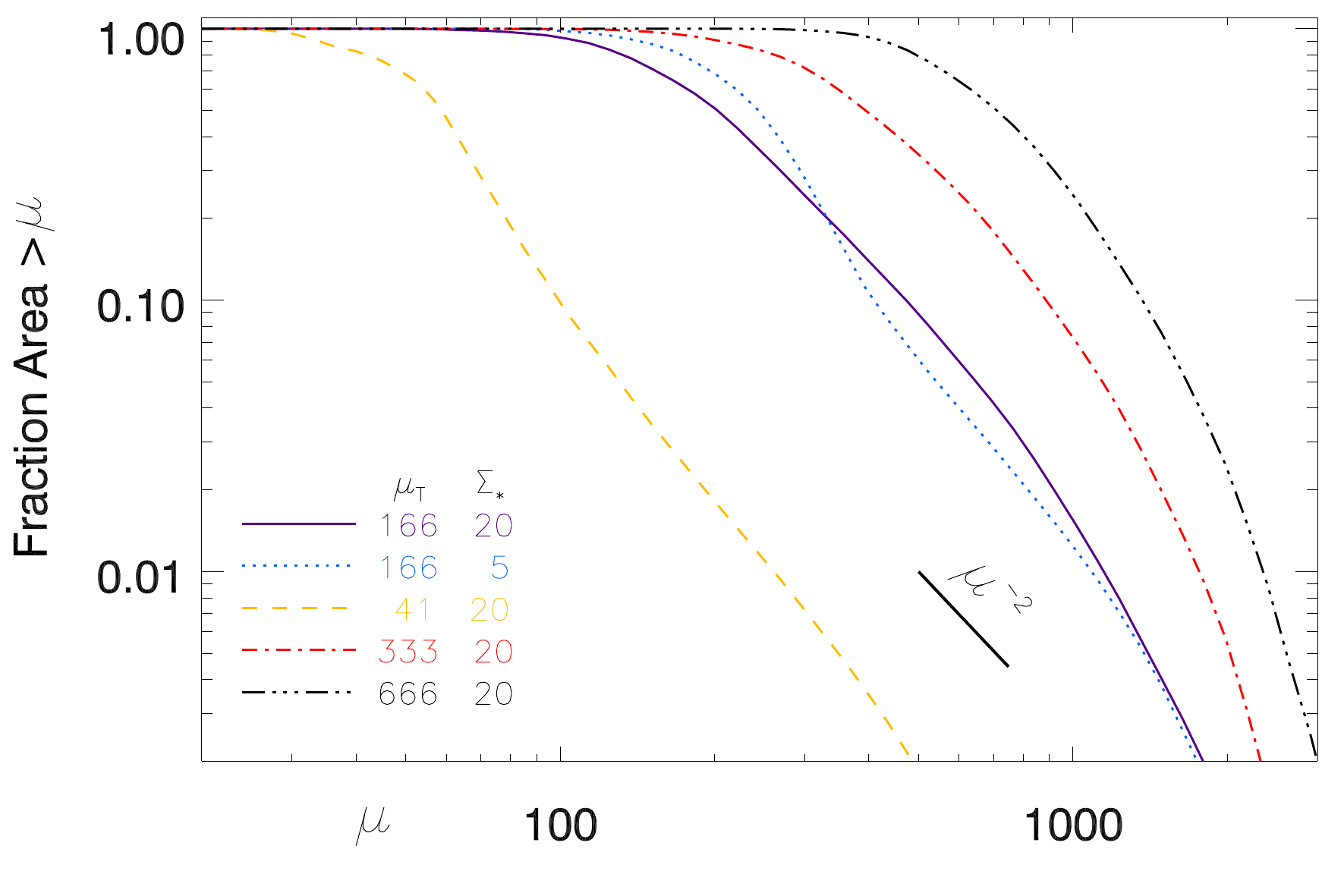} 
      \caption{Microlensing probability of magnification for the Spock galaxy. {\it Left:} Relative probability of magnification in different scenarios where the macromodel tangential magnification ($\mu_T$) or the surface mass density of microlenses ($\Sigma_*$ expressed in units of $\Msun$\, pc$^{-2}$) is varied. In all cases, the radial component of the macromodel magnification is fixed to $\mu_R=1.63$. Different values of the tangential magnification can be interpreted as different distances to the critical curve. At the lowest magnification values considered, the probability of large magnification follows the expected $\mu^{-2}$ law. For larger values of $\mu_T$, the probability resembles a log-normal. {\it Right:} Fraction of area in the source plane with magnification factors above a certain value. The models are the same as in the left panel. 
      }
         \label{Fig_ProbMu}
\end{figure*}

We make a set of simulations varying the macromodel magnification, $\mu=\mu_T\mu_R$, and the surface mass density of microlenses. The resulting probability of magnification for each simulation is presented in Figure~\ref{Fig_ProbMu}. The left panel shows the distribution of the magnification when microlenses are present. Each curve corresponds to a different scenario where the macromodel tangential magnification $\mu_T$ (which is the one that varies most along the arc) and/or the surface mass density of microlenses is varied. In all cases, the radial magnification is fixed to $\mu_R=1.63$. To first order, the most likely magnification is close to the macromodel value, while the width of the probability distribution scales with the effective surface mass density of microlenses, $\Sigma_{\rm eff} = \mu \Sigma_*$. This behaviour was observed in earlier work \citep{Diego2019}, and it is studied in more detail by \cite{Palencia2023}. For the smaller values of the magnification, the probability scales as the expected $\mu^3$ at larger magnifications (or $\mu^2$ when expressed in $log(\mu)$ intervals). For the most extreme values of the macromodel magnification, the probability resembles a log-normal,  and the median of the probability falls below the macromodel value as already described in earlier work \citep{Diego2018,Diego2019,Palencia2023}. In the right panel of Figure~\ref{Fig_ProbMu} we show the cumulative distribution for the magnification, and for the same scenarios shown in the left panel.  

 If we consider the probability of magnification to be over a factor 1000, we see from this figure that it is  above $\sim 2\%$ for any star that is within 1\,pc from the cluster caustic (or macromodel magnification greater than $166\times1.63=270$). Surprisingly, the case where the surface mass density of microlenses is $\Sigma_* = 5\, \Msun\,{\rm pc}^{-2}$ has the same probability as the case where $\Sigma_* = 20\, \Msun\, {\rm pc}^{-2}$. This unexpected behaviour is explained by the tail of the magnification of the case with $\Sigma_* = 5\, \Msun\, {\rm pc}^{-2}$, which maintains the $\mu^2$ scaling longer (i.e., toward higher magnifications) than the case with $\Sigma_* = 20\,\Msun\, {\rm pc}^{-2}$, that for $\mu\approx 1000$ is already falling faster than $\mu^2$, as shown in the left panel. For macromodel magnification $\mu = 666\times1.63=1086$, the probability of having magnification greater than $\mu=1000$ is already 20\%, but the star spends most of its time with magnification values below $\mu=800$. 

\section{The Spock galaxy}\label{Sect_Spock}
In this section, we present constraints on the physical parameters of the Spock galaxy. Relevant for this work are its size and geometry, relative position in relation to the main caustic, and stellar population derived from spectral energy distribution (SED) fitting.

\subsection{Luminosity and intrinsic size of Spock galaxy}\label{Sect_LumSpock}
The third image of the Spock galaxy has an apparent AB magnitude of 26.6 in the F110W filter. When demagnified by a factor 3.5 this corresponds to an AB magnitude of 27.96. Since the galaxy is at $z=1.0054$, the F110W filter corresponds to the visible part of the spectrum. 

  To obtain a constraint on the intrinsic size of the galaxy, we use this third image, which has a much lower magnification of $\mu=3.5$. Unlike the elongated arc which shows only a portion of the source galaxy, the third image includes the full extent of the galaxy so it gives us the full picture of the demagnified source. At the position of the third image we obtain for the tangential and radial components of the magnification $\mu_t=2.2$ and $\mu_r=1.6$, respectively. The third image can be described, to first order, as an ellipse with semimajor axis $a\approx 0.2''$ and $b\approx0.14''$. When demagnified into the source plane this corresponds to an almost circular shape with radius $r\approx 360$\,pc.

 Now we derive a separate constraint on the portion of the galaxy that is multiply imaged, but using the elongated arc instead, which contains only a portion of the Spock galaxy. By matching the magnification predicted by the caustics with the magnification of the arc observed in the image plane at the two extremes of the longest arc ($\mu\approx 14$ at each end), and using the scaling law $\mu=270/\sqrt{d\,({\rm pc})}$, we find that the edge of the Spock galaxy must be $\sim 100$\,pc away from the caustic. 
 The right panel of  Figure~\ref{Fig_SpockCaustics} shows our best possible interpretation of the geometry of the source in the caustic region. As noted above, a significant portion of the galaxy must remain outside the high-magnification region in order to not be multiply lensed. Only the small portion inside the caustic region gets imaged into the elongated arc. For our discussion, the most interesting constraint is the length of the caustic that intersects the Spock galaxy, which we estimate as $\sim 530$\,pc (see right panel of Figure~\ref{Fig_SpockCaustics}).

\subsection{Luminosity of lensed stars}
After deriving the size of the source and how the magnification scales with distance to the caustic, we can now estimate the probability of observing microlensing events, similar to the two events of \cite{Rodney2018} and the two events of \cite{Kelly2023}. First, we focus on S1 and F1, which are expected to have the largest magnifications from our lens model. In what follows, we assume that S1 and F1 are two different stars, but it is also possible that they are the same star with the critical curve passing through the midpoint between them. This distinction will not have a significant impact in the discussion below. 
Since the stars allegedly responsible for these events are only observed during short microlensing events, lasting days to weeks, they must fall below the detection limit when only the macromodel magnification is magnifying these stars. For the two lens models shown in Figure~\ref{Fig_SpockCritCurves}, the predicted magnification (without microlensing) is at least 350 for S1 and 180 for F1. 

 For the time being, let us adopt a fiducial value $\mu = 270$ for the macromodel magnification of each star, and assume F1 is a counterimage of S1; other values of $\mu$ are discussed later. Based on the scaling for the magnification, this corresponds to a distance of 0.25\,pc from the caustic (the total magnification is 540, which corresponds to a separation of 0.25\,pc from the caustic).
If the magnification at each of the two positions is 270, the star must be fainter than AB magnitude 34.6, such that in the absence of a microlensing boost in the magnification, the star remains undetected below the threshold of 28.5\,mag for a point source in the HFF. This constraint is easily satisfied by the vast majority of stars at $z=1.0054$. During a microlensing event, these stars can get a momentary boost of $\sim 1$--3\,mag, making stars as faint as AB magnitude 37 at $z=1$ detectable. The distance modulus at this redshift is 42.63\,mag, hence the star needs to be fainter than absolute magnitude $-8$ in order to remain undetected without the aid of microlensing (ignoring specific $K$-corrections, that we address in more detail below). Absolute magnitude $M_B=-8$ is the realm of bright SG and especially hypergiant stars that can approach absolute magnitudes $M_V\approx -10$ \citep{Humphreys1978}. Then, the star responsible for S1 and F1 is unlikely a hypergiant, since otherwise it would be detectable at all times in the Spock galaxy, even without the occasional microlensing boost.
More specifically, the star needs to have bolometric luminosity $L<3\times 10^7\,\Lsun$ to remain undetected in the F814W filter, or  $L<1.2\times 10^7\, \Lsun$ to be undetected in the F200LP Flashlights observations.
On the other hand, since microlensing can only boost the observed flux by 1--3 mag, the stars cannot be fainter than absolute magnitude $M_V\approx -5$; otherwise, they would be undetectable even with microlensing. Stars with this luminosity are in the red and blue SG class, so this is possibly our best candidate for S1 (and F1). 
SG stars can have a wide range of temperatures, but can reach luminosities between $\sim 10^4\, \Lsun$ and $\sim 10^5\, \Lsun$. 

Similar arguments can be followed for S2 and F2, except for this case the magnification from our lens model is almost an order of magnitude smaller ($\mu \approx 40$). Hence, without a second caustic crossing through S2--F2, and assuming S2 is a counterimage of F2, the star responsible for these two events must be 2\,mag brighter, having absolute magnitude $M_V\approx -7$. This type of star is very rare, making this possibility less likely than a second caustic crossing between S2 and F2 that would increase the magnification and consequently lower the required luminosity of the background star.


For the remainder of this work, we assume that the stars detected in the Spock arc are SGs, with temperatures ranging from $T \approx 3500\K$ to $T \approx 30,000\K$. This is in good agreement with the photometry observed for S1 and S2, and shown in Figure~\ref{Fig_Photometry}. For the cooler red SG stars, and according to the Stefan Boltzmann law, the lower temperature is compensated  by their much larger radius in such a way that $L \propto R^2T^4$ can remain very high and relatively uniform. The hotter stars are smaller in radius, but emit significantly more energy at the shortest wavelengths. Related to the wide range of possible temperatures, we need to also consider the detection band, since stars at $z\approx 1$ with $T \approx 3500\K$ will have their peak emission in the IR bands (assuming a blackbody spectrum and Wien's law), while a hot (mostly B-type) SG at the same redshift with  $T \approx 30,000\K$ will peak in the UV part of the spectrum. We then consider two {\it HST} filters, F814W in the near-IR (used for the detection of the original Spock events S1 and S2) and the bluer F200LP (used in the Flashlights program to detect F1 and F2). F814W is well suited for stars showing a strong Balmer break ($10,000\K<T<20,000\K$), while F200LP is better suited for hotter stars ($T>20,000\K$).

To compute the apparent magnitudes of the lensed stars in these filters, we need to adopt a model for their spectrum, $f_{\lambda}$. We assume the spectrum of these stars is well described (to first order) by a blackbody model with temperature $T$. We are hence ignoring spectral features such as absorption or emission lines, as well as discontinuities such as the Balmer break, but this simple approximation will still produce reasonable predictions. Setting the detection limit in the band to a given magnitude, $m_{\rm min}$, we can, for a given temperature, compute the minimum luminosity for a star at redshift $z=1.0054$,
\begin{equation}
 L_{\rm min} = \frac{L_f}{10^{0.4\Delta_{AB}}}\,, 
 \end{equation}
where $\Delta_{AB} = m_{\rm min}-m_f$ and $m_f$ is the apparent magnitude of a star at redshift $z=1.0054$ in the chosen filter and with a fiducial bolometric luminosity $L_f$ (that without loss of generality we fix to $L_f=10^7\,\Lsun$): 
\begin{equation}
m_f =  -2.5\,{\rm log}_{10}(F_f)-48.6, 
\label{eq_mab}
\end{equation}
with $F_f$ the usual flux received in the chosen filter and corrected by the filter response (and redshift dependence),
\begin{equation}
F_f(\nu) = (1+z)\frac{\int{f_f(\frac{\lambda}{1+z})S(\lambda)\lambda d\lambda}}{\int{S(\lambda)(c/\lambda) d\lambda}}, 
\end{equation}
where  $f_f(\lambda)$ is the flux density and $S(\lambda)$ is the filter and detector response \citep{Bessell2012}. The multiplicative factor $(1+z)$ accounts for the redshift correction to the flux density. This flux is given simply by
\begin{equation}
f_f(\lambda)=\frac{L_f}{4\pi D_l(z)^2}\frac{BB(\lambda,T)}{\int{BB(\lambda,T)}d\lambda}\, ,
\end{equation}
where $D_l(z)$ is the luminosity distance and $BB(\lambda,T)$ is the blackbody spectrum for temperature $T$. 
Hence, the value of $L_{\rm min}$ represents the bolometric luminosity of the star (after accounting for magnification) needed to match the chosen magnitude limit, so stars with bolometric luminosity $L$ and amplified by a factor $\mu$ are not detected if $(\mu\times L)<L_{\rm min}$.

Next, we need to find the maximum possible luminosity in the portion of the Spock galaxy that is being lensed. For this, we can use the fact that during the majority of the observations of the Spock arc, we observe no lensed star. We interpret this as the star being magnified by the most likely value during these observations, which is given by the mode of the curves shown in the left panel of Figure~\ref{Fig_ProbMu}. If a star is not observed when magnified by the mode, $\tilde{\mu}$, then its flux must be smaller than 
\begin{equation}
L_{\rm max}= 1.3\frac{L_{\rm min}}{\tilde{\mu}}\,.
\label{Eq_Lmax}
\end{equation}
The factor 1.3 in the above expression is to account for the possibility that the star can also have magnifications 30\% smaller than the mode. Setting this value to 1 has a small impact on our results, but we adopt the value of 1.3 as conservative. If $L_{\rm min}$ were interpreted as the minimum bolometric luminosity (after accounting for magnification) that a star has to have in order to be detected, $L_{\rm max}$ must be interpreted as the maximum luminosity (before accounting for magnification) that a star can have so it is not observed during the majority of the observations. The two constraints can be combined into the final condition for the bolometric luminosity, $L$, of the lensed stars:
\begin{equation}
\frac{L_{\rm min}}{\mu} < L < L_{\rm max} = 1.3\frac{L_{\rm min}}{\tilde{\mu}}\, . 
\end{equation}

Finally, we relate the bolometric luminosity with the mass of the star and the mass of the star with the IMF of the stellar population in the Spock galaxy.
For the luminosity--mass relation ($L$--$M$) of main-sequence (MS) stars, we follow \cite{Duric} with a steep $L$--$M$ for the least massive stars and an Eddington-limit type of $L$--$M$ relation for stars more massive than 55\,$\Msun$:

\begin{equation}
\frac{L}{\Lsun} = A \left ( \frac{M}{\Msun} \right )^{\beta}\, ,
\label{eq_LM}
\end{equation}
where $A =$ (0.23, 1.0, 1.4, and 3.2) $\times 10^4$ for the mass intervals $ M < 0.43\,\Msun, 0.43\,\Msun < M < 2\,\Msun, 2\,\Msun < M < 55\,\Msun$, and $55\,\Msun < M$ (respectively), and $\beta=2.3$, 4, 3.5, and 1 (respectively) for the same mass intervals. In this work, and since we require intrinsically high luminosities, only masses above $\sim 20\, \Msun$ are relevant for our calculations. 
This relation between mass and luminosity is valid for stars on the MS. Red giants and SGs or white dwarfs do not follow this relation. White dwarfs are not relevant for our study, but red giants and SGs are.  In order to relate masses with the IMF, we make the simplification that the abundance of red SGs is comparable to the abundance of hotter stars with similar luminosity. That is, given a luminosity, we derive the mass from the above relation, and from that mass we derive the abundance of stars with that mass. Finally, we assume that stars with this mass can have a range of temperatures where the only parameter is the slope ($\delta$) of the distribution (related to the blue SG to red SG ratio, or simply B/R ratio):
\begin{equation}
dN/dT \propto T^{\delta}\, .
\label{eq_dNdT}
\end{equation}
Ideally, we would instead use a luminosity function for the red SGs, but this is uncertain and dependent on  variables such as mass loss or binarity \citep{Neugent2020,Massey2023}. 
In the expression above, $\delta=0$ corresponds to a uniform mix of red and blue SG stars as a function of temperature. A value $\delta=-1$ results in $\sim 10$ times more red SGs than blue SGs, and consistent with some local measurements, while $\delta=1$ would be more representative of a very young population of luminous stars. As described below, our results depend mostly on a relatively narrow range of stars with luminosities close to (but below) the HD limit, and through the exponent $\delta$ in Eq.~\ref{eq_dNdT} we can control the relative abundances of blue and red SGs.  

As a simple parameterisation, we model the HD limit as a function of mass and temperature and then connect the mass to the luminosity through Eq.~\ref{eq_LM}. In particular, we define the HD limit as $M_{\rm HD}=40 + T/2500$. With this relation and for $T=3500\K$ the corresponding luminosity would be $L_{\rm HD}=6.4\times10^5\,\Lsun$, while for $T=30,000\K$ we get $L_{\rm HD}=1.2\times10^6\,\Lsun$. This dependence of the maximum luminosity on the temperature of the SG star is roughly consistent with the loci defined by the observations of nearby galaxies \cite{Humphreys1983,Davies2018,McDonald2022}.

Hence, the relevant variables for us are the HD limit, given by the approximation above, and the relative ratio of blue SG and red SG stars, B/R, which for simplicity we also parameterise as a function of temperature with Eq.~\ref{eq_dNdT}. The net number of stars with luminosity $L$ is then given by the number of stars with the corresponding mass $M$, according to the $L$--$M$ relation above and as predicted by the IMF. This is the strongest assumption in our model, but one that is easily rescalable (our results can be rescaled by just varying the number of SG stars in the Spock arc and controlling the B/R ratio with the parameter $\delta$). This simplification avoids the complexity of modelling an evolved population of intermediate-mass stars (such as red SGs) and reduces the number of free parameters in the model such as age, star rotation, and especially metallicity \citep{Langer1995,Wagle2020}, substituting these parameters with one single degree of freedom ($\delta$). 
For simplicity, we begin by presenting results with $\delta=0$, which is close to the B/R supergiant ratio found for some mass range and age  \citep{Schaller1992,Ekstrom2012}. The range $-1<\delta<1$ reproduces well the B/R ratio found in nearby young open clusters and galaxies, especially in their outskirts \citep{Eggenberger2002,Skillman2002,Dohm2002}, as in the case of the Spock galaxy where only the outer region of the galaxy is being multiply lensed. We discuss these alternative scenarios with varying $\delta$ values in Section~\ref{Sect_discussion}.

\begin{figure} 
   \includegraphics[width=9cm]{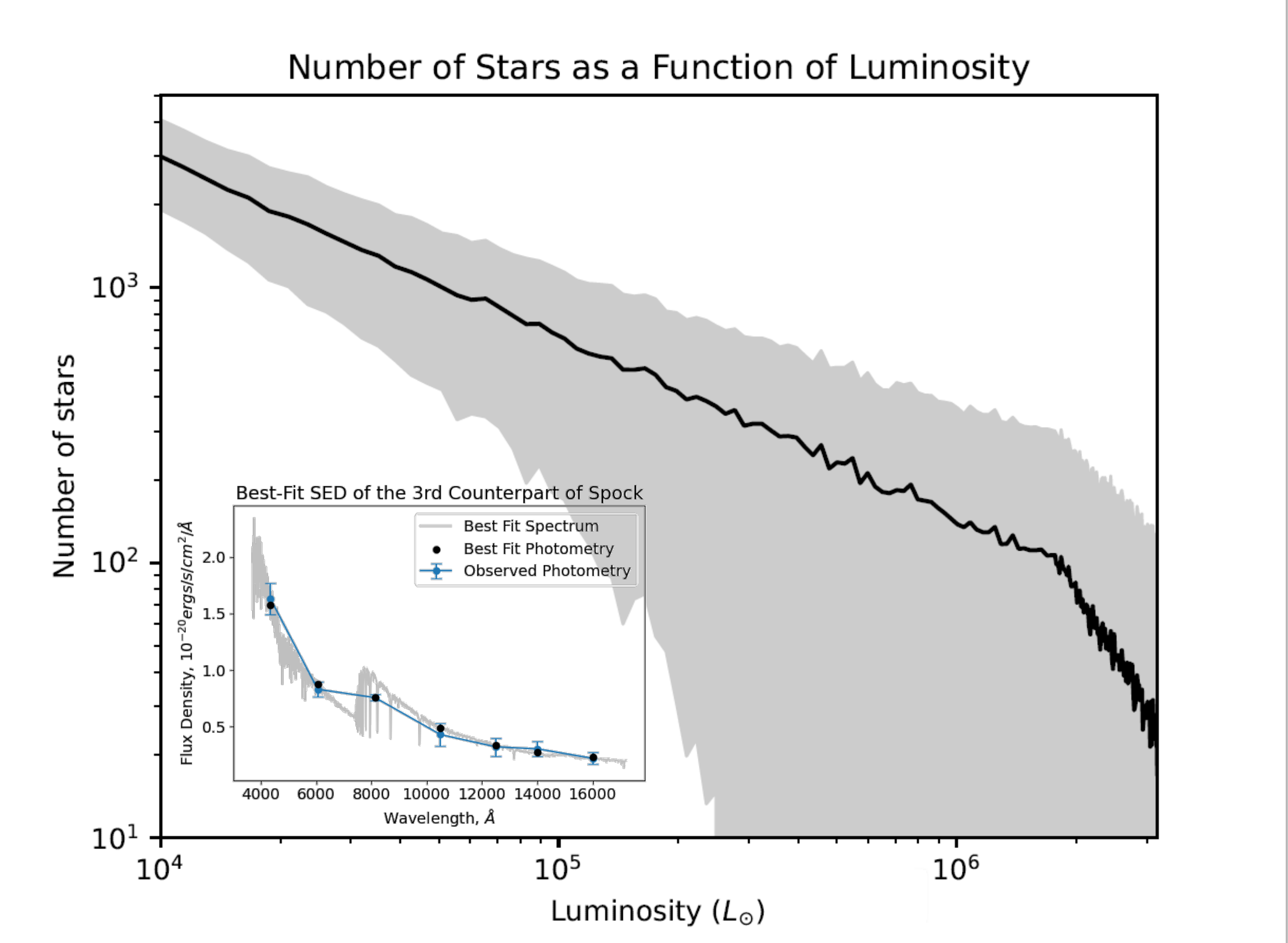} 
      \caption{Stellar luminosity function inferred from the SED fitting featured in the lower-left corner. We adopt as a fiducial model the one corresponding to the black curve in the large plot.
      }
         \label{Fig_Stellar_LumFunct}
\end{figure}

Our next step is to discuss how many of these luminous stars we expect in the small section of the Spock galaxy that is only parsecs away from the caustic.

\subsection{Abundance of giant stars near the caustic in the Spock galaxy}\label{Sect_SED}
We use the length of the Spock galaxy that intersects the caustic ($l=530$\,pc) and the scaling of magnification with distance found in a previous section, $\mu=270/\sqrt{d\,{\rm (pc)}}$, to compute the expected number of stars with a given magnification factor.  For example, the area that is being magnified with factors greater than $\mu \approx 270$ (or twice 135) is then $\sim 530\, {\rm pc}^2$. 
We can find  the number of luminous stars by fitting an IMF plus a scaling relation between mass and luminosity to the delensed AB magnitude of the Spock galaxy (27.96 in the $V$ band after correction for magnification; see Sect.~\ref{Sect_LumSpock}).


To study the stellar population in the Spock galaxy, we carry out annulus photometry on the third image to obtain the SED, as it is the only counterpart showing the entire Spock galaxy without any contamination. The circular aperture chosen is 3 times the PSF of {\it HST}, with an annulus of 1.5 times of aperture. The SED is then corrected by the lens-model magnification of $\mu = 3.5$.

We use Bagpipes \citep{Carnall2018} to carry out SED fitting, adopting a single starburst with allowed nebular emission for the fit. Bagpipes returns a young stellar population without any line emission, where the age of the Spock galaxy is $67 \pm 12$\,Myr, consistent with the blue colour observed in the images. The best-fit current stellar mass is $9.1 \pm 1.2 \times 10^{6}\,\Msun$ with a supersolar metallicity of $Z = 1.55\, Z_{\odot}$. 
These results, derived from photometry of the third counterimage, are consistent with those derived on the Spock arc by \cite{Rodney2018}, who find a stellar mass of $13.8\pm 4.5 \times 10^{6}\,\Msun$ and age $290 \pm 500$\,Myr.

After having the best-fit SED, we sample stellar populations following the Kroupa IMF, similar to what Bagpipes used. 
We find that the Spock galaxy has $2.8\times 10^4$ stars brighter than $5 \times 10^4$\,L$_{\odot}$ (see Figure~\ref{Fig_Stellar_LumFunct}). Among these, we expect $\sim 36$ stars in the 530\,pc$^2$ area considered at the beginning of this subsection, and that is magnified by factors $\mu>270$.

For the age derived above, we should not expect very massive (O-type) stars to still be present; however, given the fact that the Spock arc is magnifying a very small fraction of the entire galaxy, it is possible that a pocket of more recent star formation in that portion of the galaxy still harbors one of these massive and very hot stars. Hence, for the remainder of this work we assume that these massive stars can still exit in this portion of the galaxy.

 \section{Probability of detecting a microlensing event in the Spock galaxy}\label{Sect_Results}
Now we have all of the ingredients to do our final calculation and compute the expected number of stars that are detected above a given detection threshold and in a specific filter. 
For simplicity, we consider five regions of fixed magnification, each at a given distance from the caustic. The width of each region, $\Delta_{\rm pc}$, is determined from the scaling law $\mu=270/\sqrt{d\,{\rm (pc)}}$, by imposing that at the boundaries of the region the magnification changes by 30\% with respect to its central value. We fix $\Sigma_* = 20\,\Msun\, {\rm pc}^{-2}$ and $\mu_R=1.63$ in all regions, and vary only the tangential magnification which takes the values $\mu_T=41$, 83, 166, 333, and 666 (all shown in Figure~\ref{Fig_ProbMu}, except $\mu_T=83$, which for simplicity is omitted). Then we take the model derived from the SED fitting and compute the expected number of stars in the region of length 530\,pc and width $\Delta_{\rm pc}$ (this is an expected number, so it can be less than 1). Finally, all of the stars in this region are magnified by the corresponding constant magnification factor in the region (half the total magnification). We consider different stellar masses (or luminosities) and temperatures. For each combination of mass and temperature, we compute its bolometric luminosity using Eq.~\ref{eq_LM} and its apparent magnitude in the selected filter from Eq.~\ref{eq_mab}. If the star is brighter than $m_{\rm min}$, then it is counted as a detection.

\begin{figure} 
   \includegraphics[width=9cm]{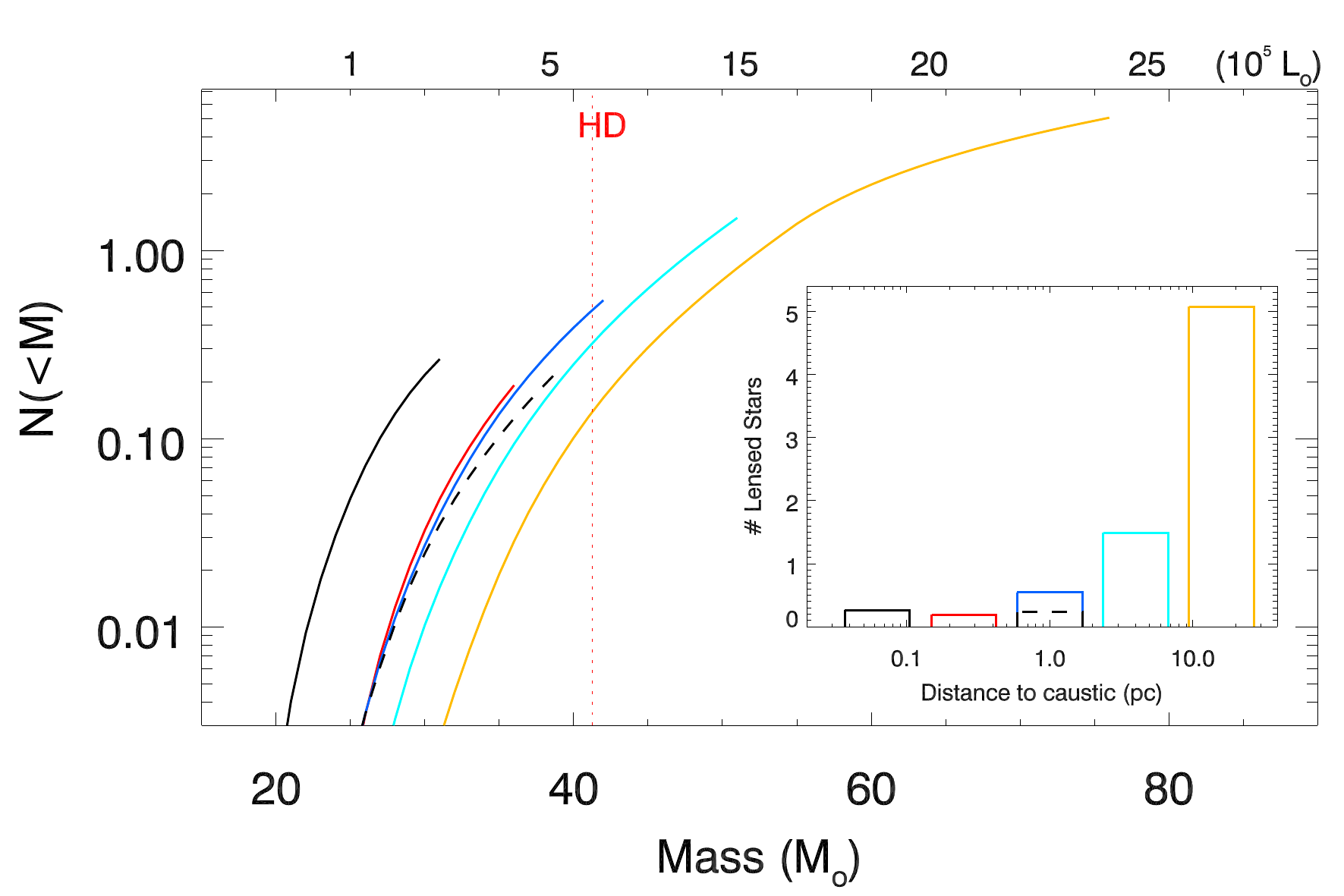} 
      \caption{
      Number of stars vs. distance. The large plot shows the cumulative number of stars observed at different distances from the caustic, as a function of the star mass. In this and other plots, the mass has to be interpreted as the equivalent mass of a MS star with the same luminosity. SG stars normally have smaller masses than its equivalent MS star. 
      We assume observations reach a limiting magnitude of 28.5 in {\it HST}'s F814W filter. Stars from the MS with masses below $\sim 20\,\Msun$ are too faint to be detected, even at the largest magnification factors considered here. Each curve stops when $L_{\rm max}$ is reached. The total number of stars expected at each distance is shown in the inset at the bottom right. Each bin in the inset is centred at a given magnification (or distance to the caustic) and the width is computed from the $270/\sqrt{d\,{\rm (pc)}}$ relation taking 30\% above and 30\% below the magnification of the bin. Colours in the inset correspond to colours in the larger plot. In all cases except for the dashed curve, the assumed surface mass density of microlenses is $\Sigma_{*}=20\,\Msun\,{\rm pc}^{-2}$, and all stars are assumed to have the same temperature of $10,000\K$. The dashed curve corresponds to separations of $\sim 1$\,pc from the caustic, but for a surface mass density of microlenses of  $\Sigma_{*}=5\,\Msun\,{\rm pc}^{-2}$. The maximum probability of seeing a lensed star is found at $\sim 16$\,pc from the caustic (yellow bin), if one ignores the Humphreys-Davidson (HD) limit (vertical dotted line). If one considers this limit, the most likely distance to observe stars is at 1\,pc (dark blue bin corresponding to total magnification $\mu \approx 270$, or $\mu=135$ for each counterimage), in good agreement with the observations and lens model. 
      }
         \label{Fig_Nstar_vs_dist}
\end{figure}

The result is shown in Figure~\ref{Fig_Nstar_vs_dist} for the five different regions, and for stars with temperature $T= 10,000\K$ being observed with the {\it HST} F814W filter and above limiting magnitude $m_{\rm min}=28.5$ in this filter. In the larger plot, we show as solid lines the cumulative number of detected stars in each one of the five distance bins, and as a function of the mass of the star (or luminosity as shown in the top scale). The curves stop where the luminosity reaches $L_{\rm max}$ (see Eq.~\ref{Eq_Lmax}).
At short distances from the caustic (black curve), the macromodel magnification is so large that faint stars can be more easily detected. The black curve also assumes that the double image of the lensed star forms an unresolved source, so its flux doubles. This situation is similar to the case of ``Earendel" or ``Godzilla," where given the large inferred magnification factors, the double image is expected to appear as a single unresolved source \citep{Welch2022,Diego2022_Godzilla}. In this situation,  the cumulative function grows faster for less-luminous stars. On the opposite extreme, at large distances from the caustic (yellowish green curve), where the macromodel magnification is smaller, only the  brightest stars can be detected. These are more rare than less-luminous stars, but since one is integrating over larger areas, the smaller probability of seeing these stars is partially compensated by the larger area. The prediction at the smaller macromodel magnification values assumes stars can exist beyond the HD limit (marked by a vertical dotted line). If the HD limit still holds at this redshift, the number of stars detected in the yellow bin would drop by a factor of $\sim 50$. In a scenario where there are no stars above $L=6\times10^5\,\Lsun$ for temperatures $T=10,000\K$, we expect the maximum probability of detecting a star at distances of $\sim 1$\,pc from the caustic (dark-blue curve), or at magnification factors $\mu\approx 270/2$ (the factor 2 accounting for the double image, each carrying half the magnification), in excellent agreement with S1 and F1. Accounting for the contribution from neighbouring distances, and after impossing the HD limit, we find that the expected number of lensed stars at the position of S1 and F1 is of order 1, in very good agreement with the observations. 

In the same plot, we also show as a dashed line the case where the surface mass density of microlenses is reduced by a factor of four --- that is, $\Sigma_{*}=5\,\Msun$. Despite having fewer microlenses, the probability of large magnification is similar to the scenario where we have four times more microlenses, as discussed earlier, resulting in a similar number of predicted lensed stars. Hence, our results are relatively insensitive to the particular choice of  $\Sigma_{*}$, as long as it is within the constrained range.

The final number of stars in each region is the result of two competing effects. On one side, larger magnification factors make the detection of the more abundant fainter stars easier, but on the other side, regions with larger magnification factors are considerably smaller. There is a sweet spot where the combination of the two effects maximises the probability of seeing a star.  In the inset plot of Fig.~\ref{Fig_Nstar_vs_dist}, we show the total number of detected stars as a function of the distance to the caustic. The width of the bins in the inset corresponds to the width $\Delta_{\rm pc}$ used in the calculation. Again, in this plot we have assumed that stars can exist beyond the HD limit. If one sets the HD limit as the maximum luminosity, the light blue and green bins would have $\sim 5$ and $\sim 50$ times fewer lensed stars, respectively.  In that case, we deduce that the best chance of finding lensed stars is when the total magnification of the double image is $100< \mu < 300$,  corresponding to a physical distance of one to a few parsecs from the caustic (dark-blue ad light-blue curves, respectively). The fact that we do not observe numerous lensed stars at larger distances from the critical curve is a direct indication that the HD limit must also be applicable at this redshift.

\begin{figure} 
   \includegraphics[width=9cm]{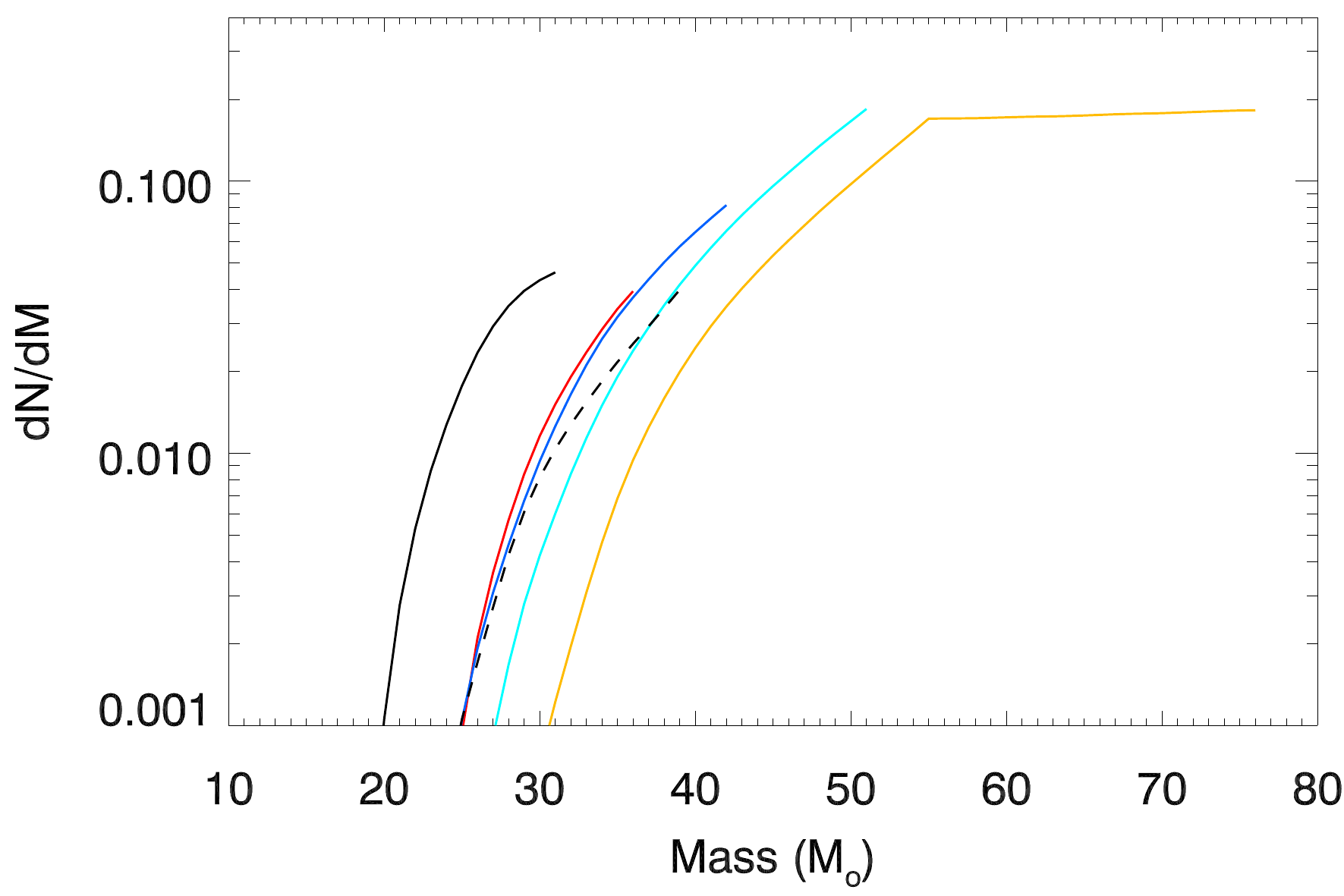} 
      \caption{
      Mass dependence vs. distance in {\it HST}'s F814W filter at depth 28.5\,mag. This figure is the differential version of Figure~\ref{Fig_Nstar_vs_dist}. The colours of the curves are the same as in the previous figure. The dashed line again represents the case where  $\Sigma_* = 5\,\Msun\,{\rm pc}^{-2}$ as in Figure~\ref{Fig_Nstar_vs_dist}. The plateau in the yellow curve marks the transition in the $L$--$M$ relation at $M\approx 55\,\Msun$. 
      }
         \label{Fig_dNstar_vs_dist}
\end{figure}

The dependence with the mass (or luminosity) of the star is better appreciated in Figure~\ref{Fig_dNstar_vs_dist}, where we show the expected number of detected stars above AB magnitude 28.5 in {\it HST}'s filter F814W, and per mass interval. At short distances from the main caustic (large magnification factors), the fraction of stars that is detected is proportional to the adopted IMF. The black solid curve corresponds to the case with maximum magnification and assuming both counterimages of the lensed star form an unresolved pair (so the observed flux doubles, making the detection of fainter stars possible). At larger distances from the main caustic, as the magnification drops, lensed events naturally select the brightest stars. The most extreme case is given by the yellow-green curve (macromodel magnification $\mu\approx 67$), where the most likely stars to be detected are the most luminous ones. This dependency with the macromodel magnification offers a unique opportunity to probe the most luminous (and rare) stars in distant galaxies, since these would be the only ones visible at relatively large distances from the critical curves of powerful lenses. In the particular case of the Spock galaxy, since we do not see the predicted abundance of lensed stars at these large distances from the critical curve, we can conclude that these stars cannot exist, as discussed above.

\begin{figure} 
   \includegraphics[width=9cm]{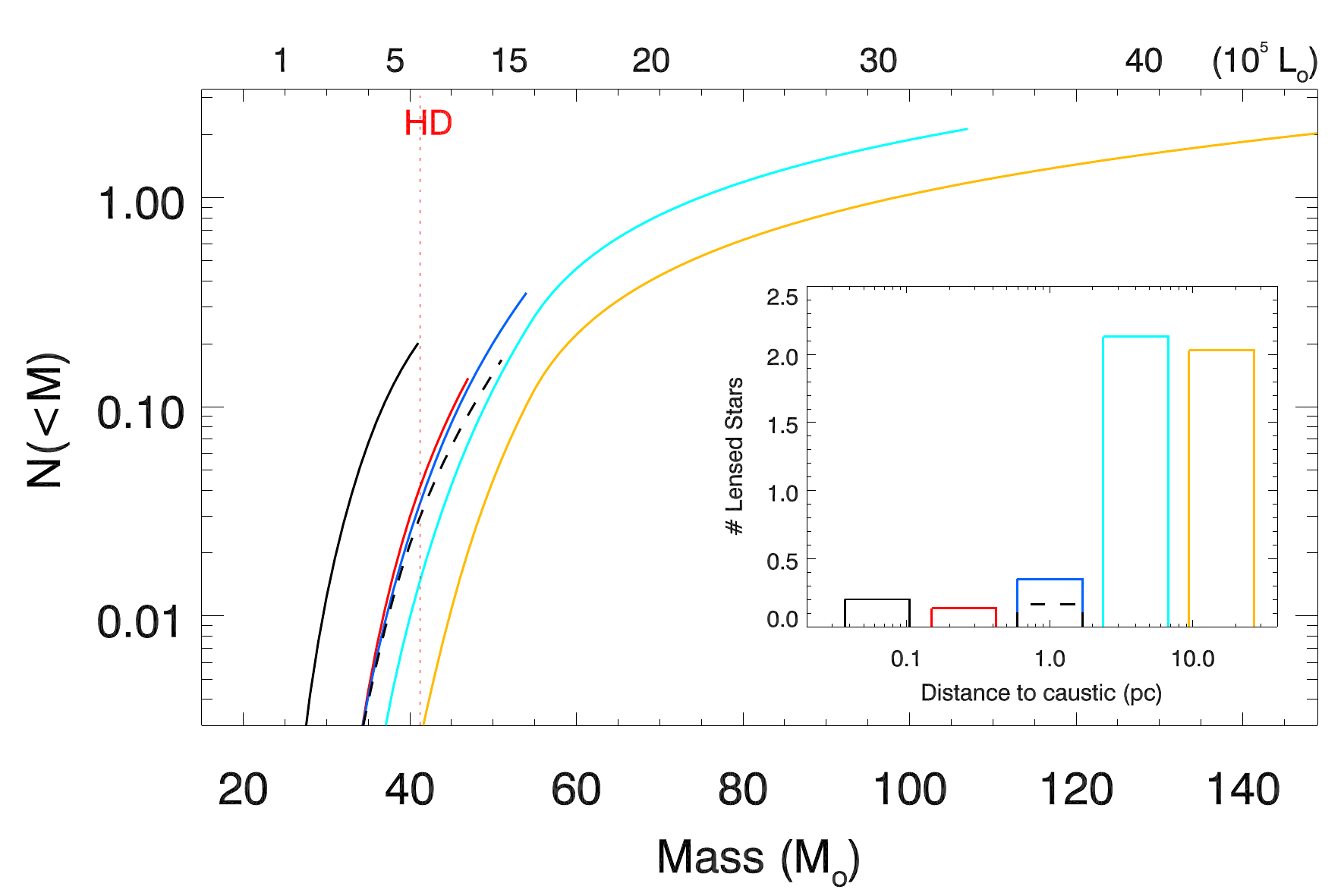} 
      \caption{Same as Figure~\ref{Fig_Nstar_vs_dist} but for a shallower observation where the limiting magnitude is 27.5 ($T=10,000\K$). Most lensed stars are predicted at larger separations from the critical curve, but if one imposes as maximum luminosity the HD limit, lensed stars are observed mostly at the largest magnification factors --- that is, closer to the caustic and critical curve.  
      }
         \label{Fig_Nstar_vs_dist_Mab27p5}
\end{figure}

We observe a general trend with the depth of the observation that can be appreciated when comparing the previous result with a shallower observation using the same filter, and assuming the same temperature for the lensed star, but reaching only 27.5\,mag. 
In Figure~\ref{Fig_Nstar_vs_dist_Mab27p5}) we have allowed $L_{\rm max}$ to be larger than the HD limit. 
This explains the counterintuitive result shown, where shallower observations predict more events in the light-blue bin than in the same bin, but for the deeper observations of Figure~\ref{Fig_Nstar_vs_dist}. This is due to the fact that $L_{\rm max}$ is larger in the shallower observation, hence the total number of lensed stars is integrated over a wider range of luminosities. That is, luminosities as high as $5\times10^6\,\Lsun$ are permitted in the calculation of the shallow case, but they are not permitted in the deeper case, which is better constrained and already excludes these luminosities because they are not observed at magnification factors $\mu=\tilde{\mu}$. As in the previous case, if the HD limit is adopted, the number of detections drops significantly and all bins contain $< 1$ star; the only exception is the black curve (closest distance to the caustic), where we expect $\sim 0.2$ lensed stars. Hence, we can conclude that in shallower observations of the Spock arc, we expect to detect lensed stars at the critical curve, or very close to it, but with a relatively low probability. In particular, we expect that one in five observations at depth F814W = 27.5\,mag may show a lensed star. But this estimate is made assuming all stars more massive than $M\approx 20\,\Msun$ have $T=10,000\K$. Cooler or much hotter stars are less likely to be detected (for the same bolometric luminosity) since the peak of their emission does not fall in the F814W filter; hence, this estimate needs to be corrected by the fraction of stars with $T\approx 10,000\K$, making the detection of such stars even less likely. 

In contrast, deep observations (as in Fig.~\ref{Fig_Nstar_vs_dist}) naturally increase the number of detections, with the bulk of them moving farther away from the critical curve. The deeper the observation, the farther one can reach down the luminosity function, and also the farther away from the critical curve one expects to see lensed stars. In this case, the larger area of smaller-magnification regions dominates over the luminosity of the brightest stars, and we predict a larger number of detections in regions where the magnification of a lensed star is $\mu\leq 150$ than in regions with larger magnification factors. For galaxies at larger redshifts, stars are fainter and one requires larger magnification factors, making the situation similar to the shallow observations at lower redshift discussed above.

\begin{figure} 
   \includegraphics[width=9cm]{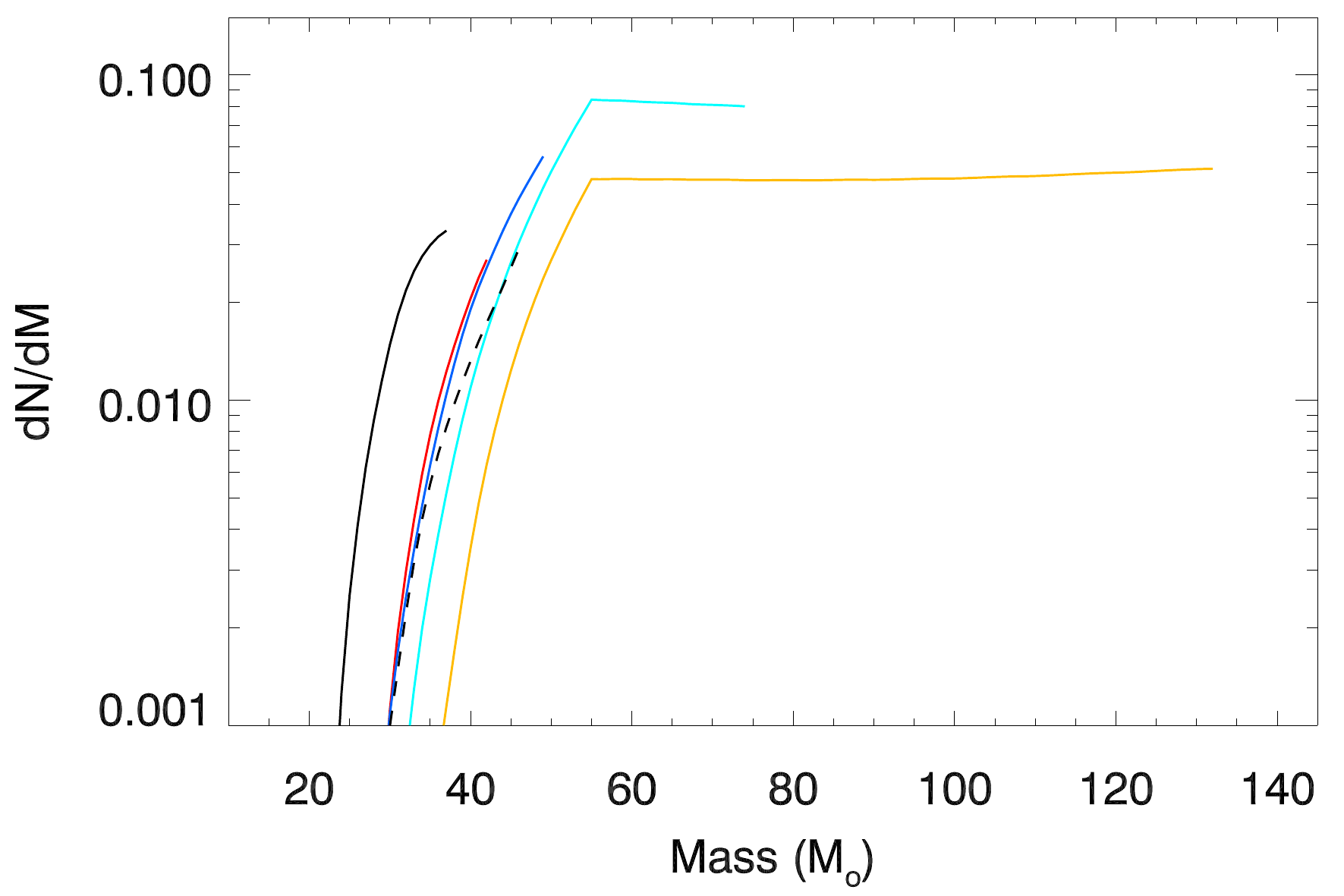} 
      \caption{
      Number of stars detected per mass interval as a function of distance to the critical curve (or magnification) in filter F200LP at depth 29\,mag (Flashlights). All stars are assumed to have the same temperature $T=10,000\K$. The colours and styles of the lines are the same as in previous figures.
      }
         \label{Fig_dNstar_vs_dist_F200LP}
\end{figure}

For an even deeper observation of the Spock arc, we considered a depth of 29\,mag similar to the one reached by the Flashlights program with the wide filter F200LP. The result is presented in Figure~\ref{Fig_dNstar_vs_dist_F200LP}. For easier comparison with the previous results, we required that all stars have the same temperature ($T=10,000\K$). 
This is a bluer filter that is more sensitive to hotter stars, so despite the increased depth, we observe a reduction in the expected number of lensed stars when compared with the shallower observations of F814W at 28.5 AB mag depth. Ignoring events above the HD limit (at $M\approx 41\,\Msun$), we expect about 40\% fewer lensed stars with $T=10,000\K$ in the Flashlights observation. However, as we see later, this filter performs much better for hotter stars, and for which the HD limit is also higher.  The range of magnifications (or distances from the critical curve or caustic) at which we expect to see lensed stars in this filter is similar to the case of the F814W filter at 28.5\,mag depth. After integrating over the entire range of distances, the number of predicted stars is $\sim 1$, similar to the case of F814W, and in good agreement with the observations.

We conclude that in general, each arc, redshift, filter, and depth needs to be studied individually since there is no general rule that applies to all. Next we study how these results depend on the effective temperature of the star.

\subsection{Dependence on star temperature}
\begin{figure} 
   \includegraphics[width=9cm]{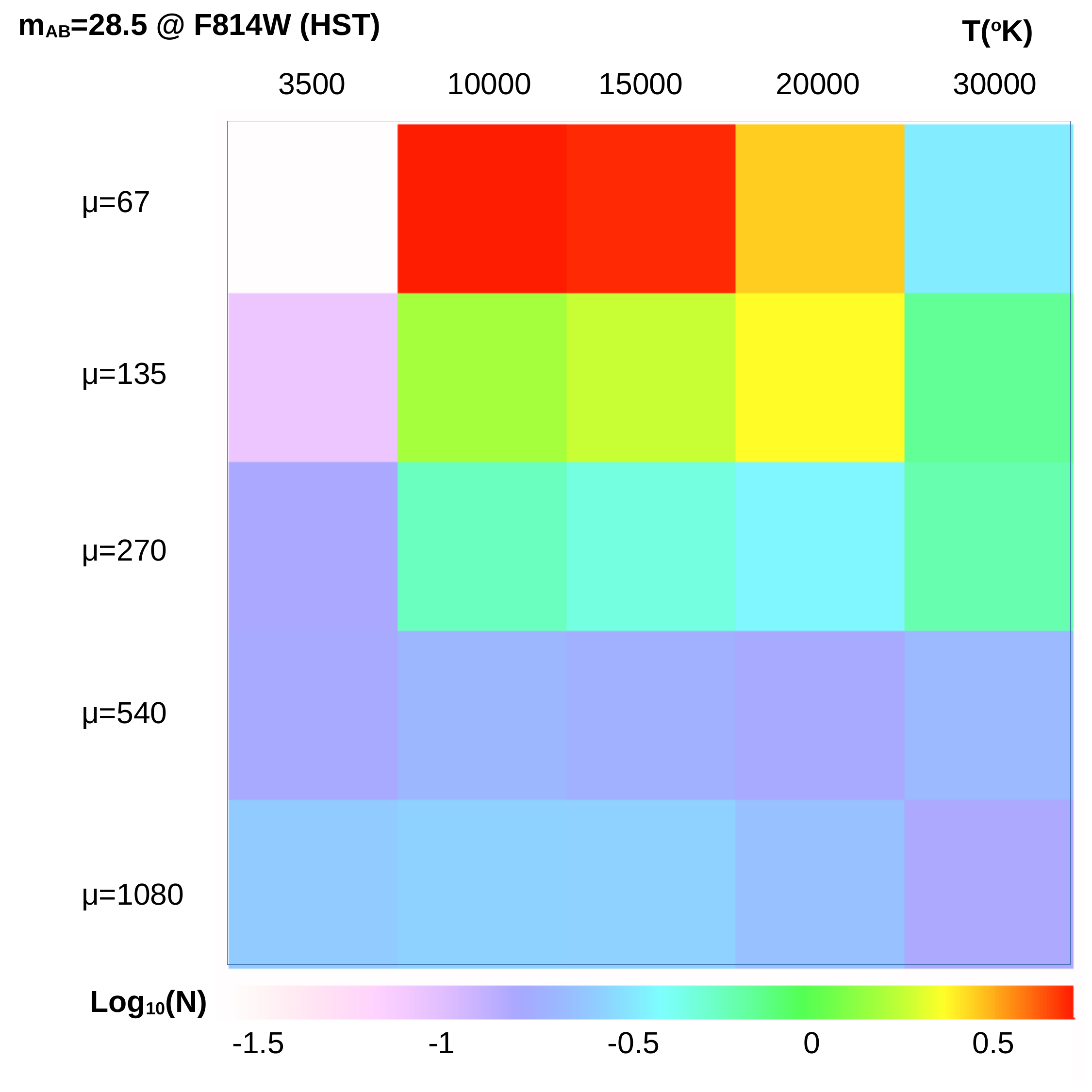} 
      \caption{
      Number of stars vs. star temperature and  macromodel magnification (or distance to the critical curve) for {\it HST}'s F814W filter and at limiting AB magnitude of 28.5. For each each bin we have assumed that all lensed stars have the temperature of that bin, so one needs to correct the total number in each bin by the fraction of stars with that particular temperature. 
      }
         \label{Fig_Nstar_T_dist_F814W}
\end{figure}

The results presented above are obtained assuming all lensed stars have the same temperature. Obviously, this is an unrealistic assumption since SG stars can have a wide range of temperatures from $\sim 3500\K$, to several tens of thousands of degrees. For a given redshift and temperature of the lensed stars, there is an optimal filter for which the star is more likely to be detected. We present the expected number of detections above AB magnitude 28.5 as a function of the star temperature for the {\it HST} filter F814W in Figure~\ref{Fig_Nstar_T_dist_F814W}. 
For this particular filter, star redshift, and depth, the most likely star to be detected has $T \approx 10,000\K$, and as discussed above, we should expect to find these stars with magnifications in the range $100< \mu < 300$. We emphasise again that at the lowest magnification values considered in this figure, only stars above the HD limit can be detected, so if these stars do not exist with the abundances we have assumed, the number of stars in these bins should be considerably reduced.

\begin{figure} 
   \includegraphics[width=9cm]{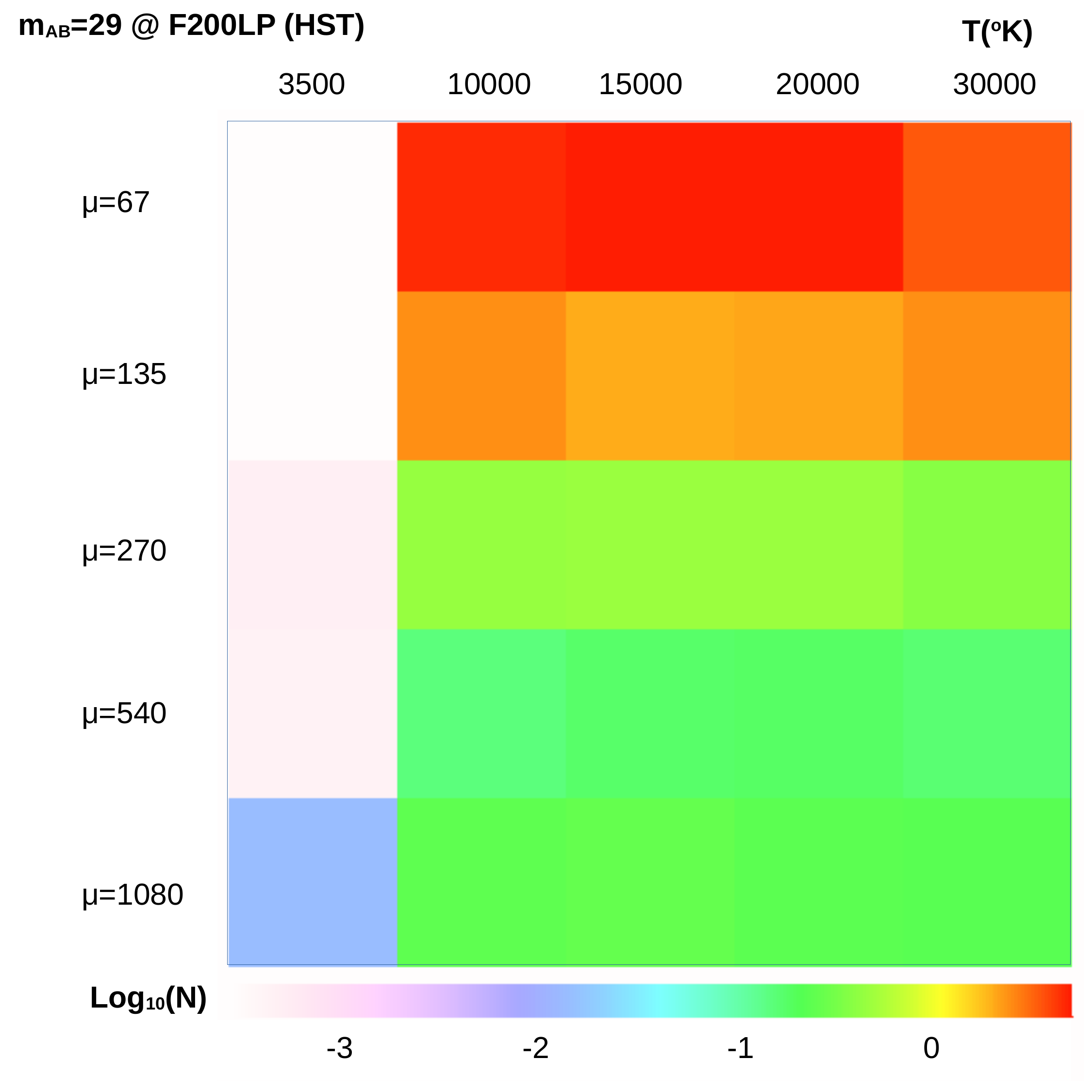} 
      \caption{
      Similar to Figure~\ref{Fig_Nstar_T_dist_F814W} but for {\it HST}'s F200LP filter and limiting AB magnitude 29. 
      }
         \label{Fig_Nstar_T_dist_F200LP}
\end{figure}

Regarding the bluer F200LP filter, the number of detections as a function of temperature and magnification is shown in  Figure~\ref{Fig_Nstar_T_dist_F200LP}). 
The number of lensed stars in the temperature range $10,000\K< T < 20,000\K$ is very similar to the shallower F814 = 28.5\,mag observation, but we observe an increase in the predicted number of events for stars with $T>20,000\K$. The reduction in number of events, when one considers the HD limit, is still applicable for the bins with the smallest magnification and $T<15,000\K$, but hotter stars can be more luminous than $6\times10^5\,\Lsun$. Hence, we expect to see more lensed stars in F200LP at 29 AB mag depth than in F814W at 28.5\,mag depth. This matches well the fact that in each of the two epochs of the Flashlights program a lensed star was spotted, while the observations with F814W had a smaller success rate. 

When we account for all of the temperatures and possible magnifications, the expected number of detected stars is summarised in Table~\ref{Tab_1}. The numbers in the columns with $\delta=0$ represent the sum of all the numbers in Figures~\ref{Fig_Nstar_T_dist_F814W} and  \ref{Fig_Nstar_T_dist_F200LP} divided by 5, which is the number of temperature bins. This factor 5 is the fraction of stars that fall in each temperature bin when one takes $\delta=0$ in Eq. \ref{eq_dNdT}. That is, we are assuming that the number of stars per temperature bin (before magnification) is homogeneous.  In the table we show the case where the HD limit is imposed and when stars above this limit are allowed to exist. The numbers in the case where the HD limit is imposed are relatively consistent with the observations (as discussed later), while the numbers when the HD limit is not imposed are not consistent. Hence, our results in Table~\ref{Tab_1} clearly favour a  scenario where the HD limit is still valid at $z=1$.
We also include the case of {\it JWST} observations at 2\,$\mu$m and reaching 29\,mag. The gain provided by {\it JWST} is obvious, providing a factor $\sim 8$ more detections. If our model is correct, in every single pointing of {\it JWST} we should observe $\sim 5$ lensed stars. Different observations will show new stars,  or variations in the flux of stars observed in previous observation, or a combination of both (new stars and flux fluctuations). The high sensitivity of {\it JWST} to the most luminous red SGs at wavelengths $\lambda > 2\,\mu$m will allow a constraint on their maximum luminosity and establish the validity of the HD limit at this redshift.
The last two columns in Table~\ref{Tab_1} show the predictions for alternatives values of the B/R ratio (or $\delta$). These are discussed in more detail in Section~\ref{Sect_discussion}.\\

\begin{table}[h!]
\caption{Number of lensed stars expected with and without imposing the HD limit. 
}\label{Tab_1}
\begin{center}
\begin{tabular}{ |c|c|c c c c| } 

\hline
   Filter    & Depth &  no HD$_{\delta=0}$     & HD$_{\delta=0}$  & HD$_{\delta=1}$ & HD$_{\delta=-1}$ \\
 \hline
 F814W  & 27.5 & 2.1 & 0.18 & 0.15 & 0.13 \\ 
 F814W  & 28.5 & 4.6 & 0.9 & 0.8 & 0.6 \\ 
 F200LP & 29   & 5.6 & 1.0 & 1.2  & 0.5 \\ 
 F200W  & 29   & 17.5 & 4.6 & 1.5 & 10.5 \\ 
 \hline
\end{tabular}
\end{center}
 \tablefoot{Numbers represent the total of the temperature--magnification diagrams in Figures~\ref{Fig_Nstar_T_dist_F814W}, \ref{Fig_Nstar_T_dist_F200LP}, and \ref{Fig_Nstar_T_dist_F200W}. The slope $\delta$ in Eq.~\ref{eq_dNdT} controls the relative abundance of red SGs vs. blue SGs. F814W and F200LP are {\it HST} filters, while F200W is a {\it JWST} filter.}
\end{table}

\subsection{Predictions for JWST}
It is natural to expect the number of discovered lensed stars at cosmological distances will increase with {\it JWST} observations. Owing to its sensitivity to longer wavelengths, {\it JWST} will extend the detections to lower-temperature stars, making possible the detection of red SGs with temperatures $T\approx 3500\K$. The first such example is Quyllur, a red SG at $z=2.1878$ setting the record for the first SG star detected by {\it JWST} beyond $z=1$ \citep{Diego2023}. Other lensed stars have been reported by {\it JWST} in recent months \citep{Chen2022,Mahler2022,Pascale2022,Meena2023}. 

\begin{figure} 
   \includegraphics[width=9cm]{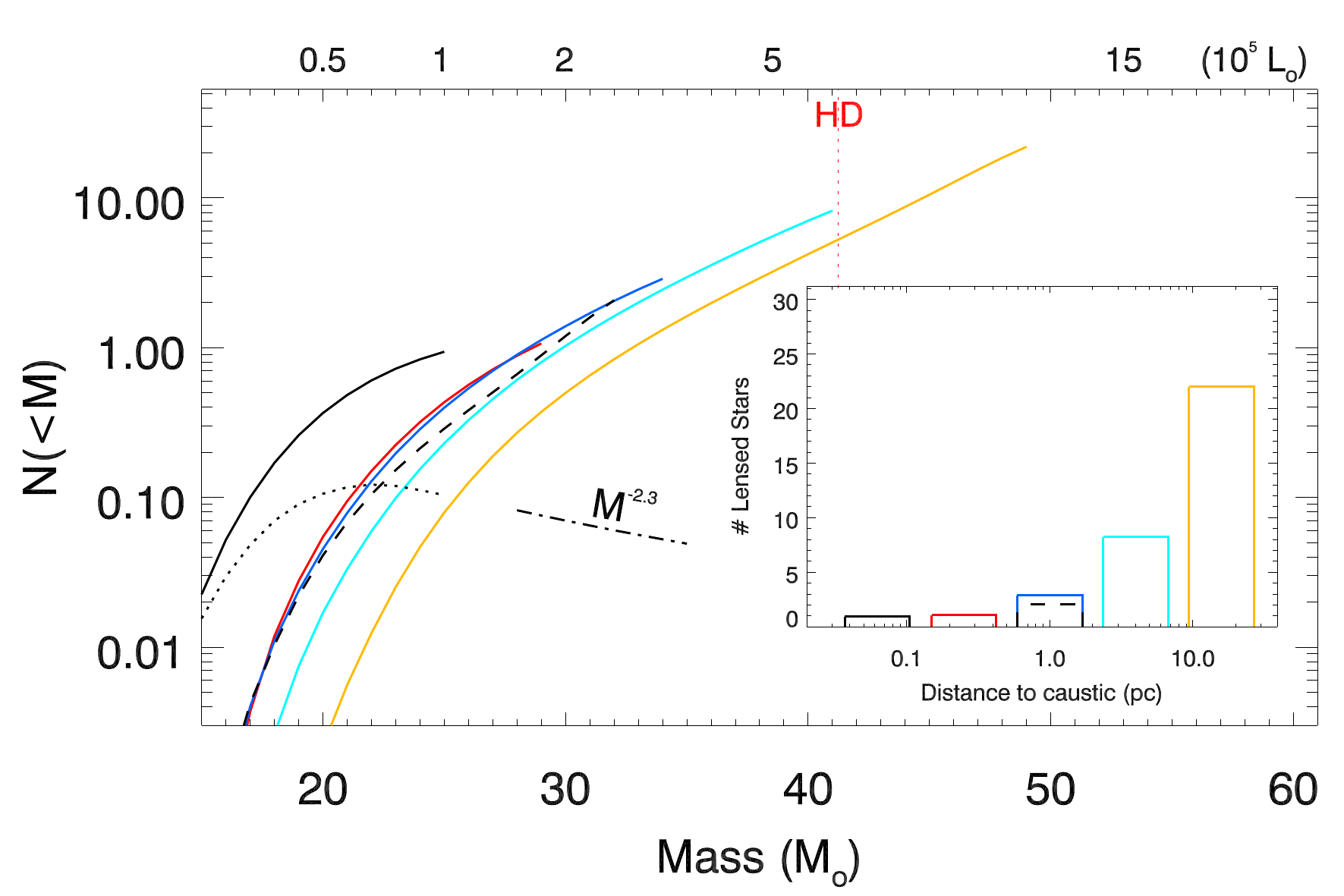} 
      \caption{
      Cumulative number of detected stars in {\it JWST}'s F200W filter, at depth 29\,mag, and for stars with $T=3500\K$. SG stars with this temperature are often less massive than implied by the abscissa.   The colours are the same as in Figure~\ref{Fig_Nstar_vs_dist}. 
      The HD limit is marked by HD and a red vertical dotted line in the plot. Cold SG stars beyond the HD limit are not observed in our local universe, but if they exist in larger numbers at $z\approx 1$, {\it JWST} has the best chance to see them in regions of the Spock arc with magnification factors $\mu\approx100$ (light-blue and yellow-green curves). The black dotted line represents the differential function $dN/dM$ of the black solid curve (corresponding to the largest magnification factor, $\mu\approx 1000$), and it shows a turnover at $\sim 20\,\Msun$.
      }
         \label{Fig_dNstar_vs_dist_JWST}
\end{figure}

First we consider an observation at 29 AB mag depth. We also consider a cold red SG star, which is more likely to be detected in the IR by {\it JWST}. The expected number of detections with {\it JWST} in the Spock arc  is shown in Figure~\ref{Fig_dNstar_vs_dist_JWST}, which corresponds to the F200W NIRCam filter and assumes one can detect stars above AB magnitude 29 in this filter. In this case, {\it JWST} will be detecting mostly red SGs, and preferably in lower-magnification regions.  At $z\approx 1$, the peak of the emission of a star with $T=3500\K$ falls in the F200W filter, so naturally this filter selects those stars at that redshift. 

One interesting result from Figure~\ref{Fig_dNstar_vs_dist_JWST} is that the differential number of detections in the bin with the largest magnification (dotted black curve) starts to drop with mass following the shape of the IMF. 
This phenomenon takes place when the minimum magnification allowed by microlensing fluctuations times the luminosity of the star is above $L_{\rm min}$. In this case, all stars more luminous than this will be detected and the number of detections traces directly the IMF. Even deeper observations with {\it JWST} will repeat the same phenomenon at larger distances from the critical curve, allowing one to probe different portions of the IMF at different separations from the critical curve.  
 Hence, {\it JWST} will detect any red SG brighter than $\sim 10^5\, \Lsun$ that is closer than 0.1\,pc from the caustic. This corresponds to an area of $\sim 50$\,pc$^2$ in the Spock galaxy; hence, a lack of detection of SGs in this regime implies a density $\rho_{*}<0.02\,{\rm pc}^{-2}$. 

\begin{figure} 
   \includegraphics[width=9cm]{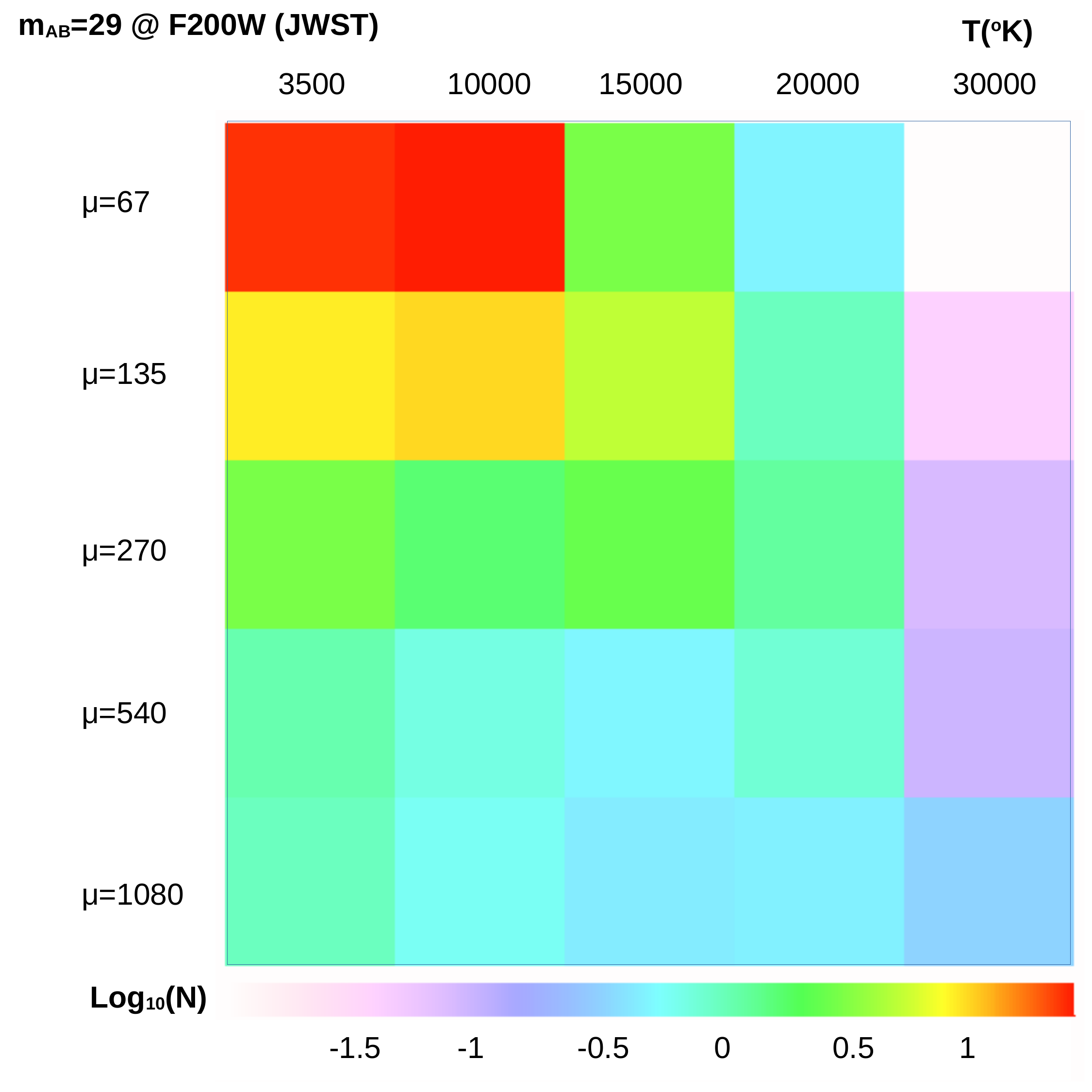} 
      \caption{
      Similar to Figure~\ref{Fig_Nstar_T_dist_F814W} but for {\it JWST}'s F200W filter and limiting AB magnitude 29. 
      }
         \label{Fig_Nstar_T_dist_F200W}
\end{figure}

When considering the dependence with temperature, as shown in Figure~\ref{Fig_Nstar_T_dist_F200W}, we predict a substantial number of red SGs to be detected by {\it JWST} at moderate magnification factors $\mu \approx 67$.  These would be spread along the Spock arc as a population of unresolved sources, and show occasional fluctuations in flux as they move across the network of microcaustics. 
However,  as in previous cases, most of the predicted detections in this bin lie above the HD limit (see  Figure~\ref{Fig_dNstar_vs_dist_JWST}), so if these stars do not exist, the number of events in the reddest bins should be reduced accordingly. Lack of a significant number of detections by {\it JWST}, or relatively low numbers of red stars spread over the arc, would set a limit on the maximum luminosity of red SG stars at $z\approx 1$, similar to the HD limit established at $z\approx 0$. On the contrary, a signifcant number of stars detected at relatively low magnification factors would imply that red SGs can be intrinsically more luminous  at $z\approx 1$ than at $z\approx 0$.

\section{Discussion}\label{Sect_discussion}

In Section 5 we presented the expected number of lensed galaxies at $z=6$ and $z=9$. Recent work based on {\it JWST} on cluster fields appear to show a deficit of galaxies beyond redshift 5 when compared with parallel or blank fields \citep{Rachana2023}. The deficit at $z=6$ can be easily understood as a depletion effect owing to the shallower than 2 slope of the luminosity function at the faint end. We predict that a similar depletion effect must exist for $z=6$ galaxies multiply lensed by MACS0416.  However, for $z=9$ (and probably also for redshifts beyond this one), we predict that $\sim 3$ times more galaxies should be found in the cluster field than in blank fields (ignoring cosmic variance). This prediction is based on recently constrained slopes for the luminosity function that find this slope to be steeper than $\alpha=2$. If the slope is instead shallower than $\alpha=2$, a depletion effect, similar to the one at $z=6$, should be taking  place. Simply counting $z>9$ galaxies around clusters, and comparing with the census of similar galaxies in blank fields, is a simple way of probing the slope of the faint end of the luminosity function. The ratio $N_{\rm cluster}/N_{\rm field}$ can then be directly related to the magnification, but also to the slope $\alpha$ \citep{Umetsu2015,Umetsu2020}.

 Regarding the microlensing events identified in the Spock galaxy, our findings are consistent with observations only if the HD limit is still valid at $z=1$. Assuming this limit is valid at this redshift, and accounting for different temperatures, we expect that $\sim 0.5$ lensed stars should be detectable per pointing with {\it HST}'s F814W filter that reaches 28.5\,mag, as shown in Table\ref{Tab_1}. 
 Focusing on the SE event (S1) reported by \cite{Rodney2018}, and on the deeper observations from their FrontierSN program, their observations cover a period of $\sim 30$ days, with a very good cadence of $\sim 1$ observation per day except for two blind periods of $\sim 3$ and $\sim 9$ days for which there are no observations. In this period one event is reported with a duration of $\sim 4$ days, although both the beginning and end of the bright phase of the event are missed since they fall in the blind periods mentioned above. We adopt 6 days as a conservative duration for the event (above the detection threshold). Hence, the success rate of events per pointing near a critical curve is $6/30=0.2$. This is only a factor a few below our expected number of 0.9 events per pointing, and is easily accommodated by small changes (consistent with the photometry) in the IMF derived from the SED model (black line in Figure~\ref{Fig_Stellar_LumFunct}), or by parsec-scale variations in the distribution of SG stars in this portion of the arc. Hence, the predicted rate of 0.9 events per observation is reasonably consistent with the observed rate of 0.2 stars detected per pointing in the FrontierSN program. We discuss later how the predicted rate matches better the observed one when modifying the B/R SG ratio. Similarly, the F1 detection in the Flashlights program corresponds to a success rate of $1/2 = 0.5$ (two observations are carried out for each cluster in the Flashlights program, and F1 was seen in only one of them), which is perfectly consistent with our estimate of 0.5 lensed stars per observation in the F200LP filter. Our findings are also consistent with the fact that S1 and F1 are found in regions where the macromodel predicts a magnification $\mu \approx 150$--300 for each counterimage, and the hypothesis that S1 and F1 are counterimages of one another, but are detected when each one is experiencing a microlensing event at different epochs.

 As discussed above, the two events S2 and F2 can be easily interpreted in a similar way, but only if a second critical curve passes near S2 and F2. Our current lens model does not produce such a second critical curve, but small variations in the lens model should easily produce the needed critical curve \citep[as already shown in][]{Rodney2018}. If no critical curve passes near S2 and F2, then these stars must be incredibly luminous in order to be detected at such low magnification factors. Their luminosities would imply the existence of stars beyond the HD limit and would create an internal inconsistency with our own predictions on S1 and F1, since if the HD limit evolves at this redshift, we would expect significantly more lensed stars than detected near S1 and F1. We consider this possibility unlikely and instead blame the lower accuracy of our lens model in this portion of the arc. Future improved lens models are needed in order to better constrain that second possible critical curve crossing. {\it JWST} data will provide further lensing constraints for the lens model.

 
  In summary, when comparing our results with observations, our predictions exceed the number of observed lensed stars if we do not set an upper limit to the luminosity of stars with temperatures $T<15,000\K$. On the other hand, our predictions are in reasonable agreement with the observations if one considers that SG stars at $z=1$ are also less luminous than the HD limit established at $z\approx 0$. Hence, our results are consistent with the hypothesis that the HD limit does not evolve between $z=0$ and $z=1$, and that the abundance of SG stars at $z=1$ is consistent with local measurements. 

In the results discussed up to this point, we have assumed $\delta=0$ in Eq.~\ref{eq_dNdT}. This can be considered as a strong assumption since it implicitly assumes that the ratio of blue SGs to red SGs (of the same luminosity) is equal to 1. For old galaxies blue SGs are expected to be very rare, while red SGs represent the evolved state of less massive (and with longer lives) intermediate-mass MS stars.  In this sense, we can consider $\delta$ as a parameterisation of the age of the galaxy. From the SED fit to the observed photometry, and the blue colour of the galaxy, we infer that the Spock galaxy is relatively young (see Section \ref{Sect_SED}), or it has sufficient star formation to support the existence of blue SGs. So our choice of $\delta=0$ is, in principle, well motivated.  
If blue SGs are more common than red SGs, then $\delta>0$. For values of $\delta=1$, the numbers for each temperature bin in Figures~\ref{Fig_Nstar_T_dist_F814W}, \ref{Fig_Nstar_T_dist_F200LP}, and \ref{Fig_Nstar_T_dist_F200W} need to be weighted not by a factor 1/5 as done earlier, but by a factor $W=\frac{T^{\delta}}{\int{T^{\delta}}} =0.04$, 0.13, 0.19, 0.25, and 0.38 for the temperature bins $T=3500\K$, 10,000\,K, 15,000\,K, 20,000,K, and 30,000\,\K, respectively. While for $\delta=-1$ the weights are 0.53, 0.19, 0.13, 0.09, and 0.06 for the same respective bins. Obviously, $W=0.2$ for all temperatures when $\delta=0$. The weights above assume all luminous stars are distributed in the five temperature bins. This is a simplification since in a real situation there may be stars with higher temperatures. In this work we are assuming the number of stars above $30,000\K$ is negligible. This approximation is appropriate if the B/R ratio is small (such as in the $\delta=-1$ case), but not so much if the $B/R>1$.  

For $\delta=1$, and after imposing the HD limit, the number of predicted detections remains more or less constant in the F814W filter with respect to the case $\delta=0$, but it increases in the bluer {\it HST}'s F200LP filter in Table~\ref{Tab_1}. For the IR {\it JWST} filter F200W, we observe a significant reduction. This filter is well suited for cold stars at $z=1$, hence is very sensitive to their abundance.  For $\delta=-1$ (more red SGs than blue SGs), the situation reverses, and we predict a large umber of detections in F200W. The current statistics based on {\it HST} data do not allow us to discriminate between different values of $\delta$, but future observations with {\it JWST} should allow us to favour one scenario over the rest.

  Interestingly, the results discussed above are to first order insensitive to changes in the micromodel. As shown in Figure~\ref{Fig_ProbMu}, considering a smaller value of $\Sigma_*$ has a minor impact on the probability of high-magnification events. The smaller $\Sigma_*=5\,\Msun {\rm pc}^{-2}$ results in a narrower PDF for the magnification, but its median value peaks at larger magnifications, and the tail of extreme magnification still follows the canonical $\mu^{-2}$ power law past $\mu=1000$. These two effects compensate the smaller abundance of microlenses, especially at large magnification values. 

  These results are derived without taking into account specific models of stellar evolution. In particular, the relatively short lifetimes of SG stars are not being considered in this work. Instead, we have assumed that these stars must exist in the Spock galaxy to be able to interpret the observations. Our results show that the abundance of these stars is well approximated by the abundance predicted by the IMF of stars with the corresponding mass to produce such luminosities (assuming the MS $L$--$M$ relation). This reasoning can be turned around to conclude that in order to have such an abundance of luminous stars, a relatively recent episode of star formation must have taken place, which is consistent with the young age derived for the Spock galaxy from SED fitting.

The results derived above assume there are no stars beyond the HD  limit. Under that assumption, we find that our  results are roughly consistent with the observations (that is, $\sim 0.5$ observed stars per pointing). Given the relatively small area that is being amplified in the Spock arc, we also consider the possibility that stars beyond the HD limit may exist, but their surface number density is smaller than what was assumed in our previous calculations. Otherwise these stars would have been detected by the FrontierSN or Flashlights programs, preferentially at larger distances from the critical curve. Since the observed stars appear near the expected position of the critical curves (true at least for S1 and F1, and likely true also for S2 and F2, assuming a second critical curve passes between them), we can conclude that if stars with luminosities beyond the HD limit exist in this small portion of the Spock galaxy, their number density must be smaller than what was assumed in our calculations above.\\
The strongest constraint on the abundance of stars beyond the HD limit comes from the regions with smallest magnification, as shown in Figures~\ref{Fig_dNstar_vs_dist} and \ref{Fig_dNstar_vs_dist_F200LP}. Our model assumed a surface number density of $\approx 8000$ stars per kpc$^2$ with luminosities beyond $6\times10^5\Lsun$.  
From the results in Table~\ref{Tab_1}, and adopting the case $\delta=0$, the number of expected observed stars per pointing, and beyond the HD limit is simply the difference between Column 3 (labeled "No HD$_{\delta=0}$"), and Columns 4, 5, and 6 --- that is, $\sim 3.5$ for the FrontierSN program or $\sim 4.5$ for the Flashlights program. These numbers need to be reduced by approximately a factor 20 for these stars to be rare enough to not contribute significantly to the number of detections farther away from the critical curve. This translates into a surface number density of $< 400$ stars\,kpc$^{-2}$ with luminosities beyond the HD limit.
If we consider the bins with magnifications 67 and 135, where these very bright stars are more likely to be found, the area in the source plane with magnification in this range is $A\approx530\times20\approx0.01$\,kpc$^2$. This means the number of stars beyond the HD limit in this region can be up to $\sim 4$. These stars remain in general undetected without the aid of microlensing provided they are less luminous than $L_{\rm bol}<3\times 10^7\,\Lsun$, as estimated earlier.
Similar arguments can be used, but this time applied to the smaller area associated with the larger-magnification bins, and assuming that the stars responsible for the observed events are less luminous than the HD limit. Our model contains $\approx 9000$ stars\,kpc$^{-2}$  with bolometric luminosities between $\sim 10^5\,\Lsun$ and $\sim 6\times10^5\,\Lsun$. As discussed above, this number density ( equivalent to $\sim 1$ SG star per $10\times10$ pc$^2$) reproduces the observations from the FrontierSN and Flashlights programs reasonably well. Although high, this number density is consistent with the fact that SG stars are often found in large concentrations of stars \citep{Davies2007}.

 The abundance of SG stars in the Spock galaxy can be better constrained with future observations of this arc. If we consider the farthest bin in Figure~\ref{Fig_Nstar_vs_dist} at $d\approx 16$\,pc from the caustic (yellow-green bin, or total magnification $\mu\approx 70$, equivalent to regions in the image plane where $\mu \approx 35$ for each counterimage), this bin is highly sensitive to the presence of the most luminous stars (as shown more clearly in Figure~\ref{Fig_dNstar_vs_dist}). By simply mapping how far from the critical curve lensed stars are observed, one can constrain the abundance of the most massive and luminous stars, since at relatively large distances from the caustic, only the most luminous stars are detectable.

Future {\it JWST} observations will improve the statistics on the number of detected lensed stars, offering us a privileged view into the massive end of the IMF of the Spock galaxy. 
Once enough SG stars are detected by {\it JWST}, an interesting application is to reverse engineer the tip of the red giant branch (TRGB) distance indicator and use it instead to constrain the magnification. If the TRGB holds at this redshift, we can use the observed luminosities of the brightest red stars to infer their true unlensed luminosity. This gives a direct constraint on the magnification which can be used to improve on the lens models. Current models typically disagree by a factor of 2--3 in the magnification near the critical curves. If the maximum bolometric luminosity for red SGs is still $\sim 6\times10^5\,\Lsun$, {\it JWST} will be able to see these bright stars and use them as distance indicators to constrain $\mu$. There is of course the issue of microlensing that may introduce uncertainty in the magnification, but monitoring these stars should provide accurate measurements of the mode of the magnification. These stars can also show variability, although this is expected to be different than the $1/\sqrt{t}$ variability predicted by lensing (where $t$ is time to the maximum magnification). Frequent monitoring of these stars should easily distinguish between intrinsic variability and a microlensing event. Since {\it JWST} will be able to see red SGs farther away from the critical curve, at these separations the mode of the magnification is a good tracer of the macromodel magnification at that particular position. 

Another interesting property that can be checked with future observations is to estimate the presence of luminous blue companions to red SGs.  From local measurements, $\sim 20$\% of the red SGs are expected to have such companions \citep{Neugent2020b}. Monitoring of microcaustic events can unambiguously reveal companion stars provided that this companion is bright enough so it can be observed when the companion crosses a microcaustic. These events would show distinctive chromatic features since the stars in the pair would have different temperatures (and colours) which manifest themselves when each star is crossing a microcaustic.


\section{Conclusions}\label{Sect_conclusions}
We have presented an updated model of the cluster MACS0416, based on the hybrid algorithm WSLAP+, taking advantage of new weak-lensing measurements over the extended region covered by the BUFFALO program, as well as new spectroscopically confirmed multiply lensed galaxies confirmed by MUSE data. 
Our new lens model contains $4.9 \times 10^{14}\, \Msun$ in the central 1\,Mpc. The mass adopts the same bimodal distribution found in earlier work, but we find the mass contained in each substructure to be almost identical, suggesting a mass ratio 1:1 for the structures in this cluster.  

Using this lens model we predict the number of expected multiply lensed galaxies to be detected with {\it JWST} at $z=6$ and $z=9$, using the latest luminosity functions at high redshift derived from recent work based on {\it JWST} observations. For $z=6$ we predict the usual depletion effect where, despite the increased effective depth provided by lensing, we expect fewer galaxies near the cluster centre than in the parallel field, where the magnification is $\mu\approx 1$. However, at $z=9$ the situation is reversed and we predict a larger number of $z=9$ galaxies detected in the cluster field than in the parallel field. This is a direct consequence of the steeper luminosity function at $z=9$ than at $z=6$ in the faint end.

We focus our attention on four transients identified in one of the strongly lensed  arcs in MACS0416, dubbed the Spock arc. Although other interpretations are still possible for these transients \citep{Rodney2018,Kelly2023}, we demonstrate that they are consistent with the hypothesis that they are due to microlensing events of SG stars at $z=1.0054$. These microlensing events are expected to be frequent in this arc. We consider two of the {\it HST} filters (and depths) matching the real observations, the red filter F814W (and observations at 28.5 AB mag depth) and the wider, but bluer, filter F200LP (and depth of 29\,mag). From the lack of detection of lensed stars in most observations of this arc, we set an upper limit to the luminosity of detectable stars, ruling out the existence of SG stars with $L>3\times 10^6\,\Lsun$ in the portion of the arc that is being multiply lensed. 
We assume the stars that are being lensed all belong to the SG type, with luminosities at least $\sim10^5\Lsun$. These stars can have a wide range of temperatures. We consider temperatures from $T=3500\K$ to $T=30,000\K$. The abundance of these stars is originally constrained by a model based on the Kroupa IMF that we fit to the observed photometry in different bands using BAGPIPES. 
We find that the predicted number of observed lensed stars matches the observations if the HD limit has no significant evolution between $z=0$ and $z=1$. 
Alternatively, stars beyond the HD limit may still exist in this portion of the galaxy, but their number density must be $< 400$ stars\,kpc$^{-2}$ in order for their probability of being observed to be sufficiently small. On the other hand, if the stars responsible for the observed events are less luminous than the HD limit but with bolometric luminosities greater than $10^5\,\Lsun$, their surface number density must be $\sim 9000$ SG stars\,kpc$^{-2}$. 

The four observed events can be fully described as microlensing events of SG stars. In particular, the NW S2 event is consistent with  mcirolensing of a blue SG star, while the SE event is consistent with microlensing of a red SG star. For F1 and F2, the lack of colour information does not allow us to constrain the type of star being lensed, but we remind that this filter is better suited for hot stars than for red stars.
 The HD limit can be also connected to metallicity. We find that the metallicity of the Spock galaxy is likely supersolar. In this scenario, massive stars would lose mass more efficiently through radiation-driven stellar wind, tightening the HD limit.  
 
An alternative way of testing for the HD limit is by measuring the separation from the microlensed events to the critical curve. Stars with luminosities beyond the HD limit can be found at larger separations since they require smaller magnification factors in order to be detected. The Spock arc is not ideal for this purpose because there is still significant uncertainty on the position of the critical curve (or even number of them) that is (are) crossing the arc, but under the assumption that at least one of them (between S1 and F1) is well reproduced by our lens model, we also find good consistency between the observed distance to the critical curve of the microlensed events and the existence of an HD limit. 

We also make predictions for future {\it JWST} observations of this cluster. We find that in the F200W filter ($\sim 2\,\mu$m), {\it JWST} will detect $\sim 5$ SG stars per observation spread along the Spock arc. 

We have shown how there is a high variability in the number of expected detected stars depending on the depth of the observation (or redshift of the background source), filter used to perform the observation,  temperature of the lensed stars, and obviously the luminosity of the stars and their age. Each particular arc and observation needs to be modelled specifically. Other arcs in the same cluster (as well as in other clusters) have similar events  \citep[see][for events discovered as part of the Flashlights program]{Kelly2023, Meena2022}. 

\begin{acknowledgements}
 The authors wish to thank Roberta Humphreys for useful comments and feedback. J.M.D. acknowledges the support of project PGC2018-101814-B-100 (MCIU/AEI/MINECO/FEDER, UE) Ministerio de Ciencia, Investigaci\'on y Universidades.  This project was funded by the Agencia Estatal de Investigaci\'on, Unidad de Excelencia Mar\'ia de Maeztu, ref. MDM-2017-0765. 
 A.N., M.J., and DJ.L. are supported by the United Kingdom Research and Innovation (UKRI) Future Leaders Fellowship (FLF), ``Using Cosmic Beasts to Uncover the Nature of Dark Matter" (grant MR/S017216/1). 
 D.J.L. is also supported by STFC grants ST/T000244/1 and ST/W002612/1.
 M.L. acknowledges CNRS and CNES.
 P.K. is supported by NSF grant AST-1908823, and NASA/STScI grants GO-15936 and GO-16278.
 A.Z. acknowledges support by grant 2020750 from the United States-Israel Binational Science Foundation (BSF) and grant 2109066 from the United States National Science Foundation (NSF), and by the Ministry of Science \& Technology, Israel.
 A.V.F. is grateful
for financial support from the Christopher R.
Redlich Fund and numerous individual donors.
S.K.L and J.L. are supported by the Collaborative Research Fund under grant C6017-20G, which is issued by Research Grants Council of Hong Kong S.A.R. 
This research is based on observations made with the NASA/ESA {\it Hubble Space Telescope} obtained
from the Space Telescope Science Institute (STScI), which is operated by the Association of Universities for Research in Astronomy, Inc., under NASA contract NAS5–26555. These observations are associated with program(s) ID GO-15117 (P.I C. Steinhardt).
 
\end{acknowledgements}

\bibliographystyle{aa} 
\bibliography{MyBiblio} 


\begin{appendix}

\section{Arc positions and redshifts}
Strong lensing constraints used to derive the lens model. ID$_2$ and $z_2$ are the IDs and redshifts in \cite{Bergamini2022}. In general, all redshifts agree to the subpercent level, except for our system (ID = 42) where \cite{Bergamini2022} derives a redshift 14.5\% larger.
Arcs marked with $\dag$ are from \cite{Bergamini2022}, and are not included in our set of constraints since they were published after the model was derived. Arcs that have no ID$_2$ nor $z_2$ in the table are systems identified with BUFFALO and MUSE data that are included in our model, but are not included by \cite{Bergamini2022}. These are systems ID = 33, 38, 40, 43, 44, 46, 50, 53, 54, 55, 70, and 72. 

\begin{table}
  \begin{minipage}{165mm}                                               
    \caption{Full strong lensing dataset. }
 \label{tab_arcs}
 \begin{tabular}{|cccccc|}   
 \hline
  ID   &       RA    &     DEC      & $z_{\rm MUSE}$ & ID$_2$ & $z_2$ \\
\hline
1      &  64.0408675 & -24.0616447  &    1.896   & 5.1c   & 1.893 \\
1      &  64.0435396 & -24.0635339  &     -      & 5.1b   & 1.893 \\
1      &  64.0474367 & -24.0686997  &     -      & 5.1a   & 1.893 \\
2      &  64.0411592 & -24.0618356  &    1.895   & 5.2c   & 1.893 \\
2      &  64.0431533 & -24.0631203  &     -      & 5.2b   & 1.893 \\
2      &  64.0475325 & -24.0688519  &     -      & 5.4a   & 1.893 \\
3      &  64.0308421 & -24.0671572  &    1.9894  & 15.1c  & 1.990 \\
3      &  64.0353283 & -24.0710203  &     -      & 15.1b  & 1.990 \\
3      &  64.0418754 & -24.0757592  &     -      & 15.1a  & 1.990 \\
4      &  64.0308492 & -24.0672517  &    1.99    & 15.2c  & 1.990 \\
4      &  64.0352429 & -24.0710325  &     -      & 15.2b  & 1.990 \\
4      &  64.0419121 & -24.0758594  &     -      & 15.2a  & 1.990 \\
5      &  64.0324804 & -24.0684442  &    2.0948  & 16.1c  & 2.095 \\
5      &  64.0327321 & -24.0686961  &     -      & 16.1b  & 2.095 \\
5      &  64.0335933 & -24.0694772  &     -      & 16.1a  & 2.095 \\
5      &  64.0436363 & -24.0769933  &     -      &  -     &     \\
6      &  64.0398687 & -24.0631192  &    2.0881  & 7c     & 2.085 \\
6      &  64.0407412 & -24.0636225  &     -      & 7b     & 2.085 \\
6      &  64.0471708 & -24.0711372  &     -      & 7a     & 2.085 \\
7      &  64.0261300 & -24.0772783  &    2.2982  & 29c    & 2.298 \\
7      &  64.0284838 & -24.0797714  &     -      & 29b    & 2.298 \\
7      &  64.0367787 & -24.0839397  &     -      & 29a    & 2.298 \\
8      &  64.0393337 & -24.0704250  &    1.0054  & 13a    & 1.005 \\
8      &  64.0383658 & -24.0697531  &     -      & 13b    & 1.005 \\
8      &  64.0343062 & -24.0660503  &     -      & 13c    & 1.005 \\
9      &  64.0276462 & -24.0727036  &    3.2175  & 20.1c  & 3.222 \\
9      &  64.0322367 & -24.0751339  &     -      & 20.1b  & 3.222 \\
9      &  64.0404263 & -24.0815053  &     -      & 20.1a  & 3.222 \\
10     &  64.0263242 & -24.0743661  &    1.6333  & 24c    & 1.637 \\
10     &  64.0311150 & -24.0789806  &     -      & 24b    & 1.637 \\
10     &  64.0359133 & -24.0813494  &     -      & 24a    & 1.637 \\
11     &  64.0241529 & -24.0809203  &    1.9656  & 36c    & 1.964 \\
11     &  64.0284129 & -24.0845719  &     -      & 36b    & 1.964 \\
11     &  64.0316896 & -24.0857892  &     -      & 36a    & 1.964 \\
12     &  64.0235546 & -24.0817486  &    2.2182  & 37c    & 2.218 \\
12     &  64.0286679 & -24.0860025  &     -      & 37b    & 2.218 \\
12     &  64.0298979 & -24.0863944  &     -      & 37a    & 2.218 \\
13     &  64.0344096 & -24.0637628  &    2.0910  & 10c    & 2.093 \\
13     &  64.0396529 & -24.0666592  &     -      & 10b    & 2.0943 \\
13     &  64.0446321 & -24.0721236  &     -      & 10a    & 2.093 \\
14     &  64.0465296 & -24.0604303  &    3.2355  & 1c     & 3.238 \\
14     &  64.0470354 & -24.0608272  &     -      & 1b     & 3.238 \\
14     &  64.0491533 & -24.0628947  &     -      & 1a     & 3.238 \\
15     &  64.0422746 & -24.0605783  &    2.1067  & 4c     & 2.107 \\
15     &  64.0475379 & -24.0660783  &     -      & 4b     & 2.107 \\
15     &  64.0481908 & -24.0669914  &     -      & 4a     & 2.107 \\
16     &  64.0365775 & -24.0670578  &    0.9397  & 12.1c  & 0.940 \\
16     &  64.0369150 & -24.0674903  &     -      & 12.1b  & 0.940 \\
16     &  64.0409596 & -24.0712164  &     -      &  -     &  \\
17     &  64.0230608 & -24.0772986  &    5.3659  & 32c    & 5.365 \\
17     &  64.0284821 & -24.0830294  &     -      & 32b    & 5.365 \\
17     &  64.0351254 & -24.0855331  &     -      & 32a    & 5.365 \\
18     &  64.0293633 & -24.0733550  &    5.1060  & 21.1c  & 5.106 \\
18     &  64.0308662 & -24.0742067  &     -      & 21.1b  & 5.106 \\
19     &  64.0250233 & -24.0750497  &    3.4909  & 27c    & 3.492 \\
19     &  64.0294879 & -24.0799181  &     -      & 27b    & 3.492 \\
19     &  64.0375504 & -24.0836828  &     -      & 27a    & 3.492 \\
20     &  64.0227721 & -24.0746250  &    3.4406  & 30c    & 3.440 \\
20     &  64.0313208 & -24.0819283  &     -      & 30b    & 3.440 \\
20     &  64.0337146 & -24.0832186  &     -      & 30a    & 3.440 \\
 \hline
 \end{tabular}  
 \end{minipage}
\end{table}

\setcounter{table}{0}
 \begin{table}
    \begin{minipage}{165mm}                                               
    \caption{cont.}
\begin{tabular}{|cccccc|}   
 \hline
  ID   &       RA    &     DEC      & $z_{\rm MUSE}$ & ID$_2$ & $z_2$ \\
\hline
21     &  64.0416350 & -24.0600322  &    3.2885  & 3c     & 3.290 \\
21     &  64.0453321 & -24.0627914  &     -      & 3b     & 2.290 \\
21     &  64.0493079 & -24.0682117  &     -      & 3a     & 3.290 \\
22\dag &  64.0383500 & -24.0841261  &    3.253   & 28a    & 3.253 \\
22     &  64.0264042 & -24.0767289  &    3.2526  & 28c    & 3.253 \\
22     &  64.0284067 & -24.0790322  &     -      & 28b    & 3.253 \\
23     &  64.0234888 & -24.0761606  &    4.1218  & 31c    & 4.122 \\
23     &  64.0293129 & -24.0818383  &     -      & 31b    & 4.122 \\
23     &  64.0355646 & -24.0847072  &     -      & 31a    & 4.122 \\
24     &  64.0269333 & -24.0699764  &    3.8696  & -      &  \\
24     &  64.0340133 & -24.0745969  &    3.8710  & 18b    & 3.871 \\
24     &  64.0402579 & -24.0798950  &     -      & 18a    & 3.871 \\
25     &  64.0266083 & -24.0705319  &    4.1032  & 19.1c  & 4.103 \\
25     &  64.0337633 & -24.0748514  &     -      & 19.1b  & 4.103 \\
25     &  64.0401517 & -24.0803403  &     -      & 19.1a  & 4.103 \\
26\dag &  64.0468408 & -24.0753850  &    3.292   & 11a    & 3.292 \\
26     &  64.0353038 & -24.0647647  &    3.2922  & 11c    & 3.292 \\
26     &  64.0385875 & -24.0659972  &     -      & 11b    & 3.292 \\
27     &  64.0252696 & -24.0736092  &    3.0773  & 26c    & 3.081 \\
27     &  64.0305596 & -24.0792475  &     -      & 26b    & 3.081 \\
27     &  64.0377979 & -24.0824158  &     -      & 26a    & 3.081 \\
28     &  64.0255200 & -24.0736789  &    3.1103  & 25c    & 3.111 \\
28     &  64.0304467 & -24.0790472  &     -      & 25b    & 3.111 \\
28     &  64.0381496 & -24.0824269  &     -      & 25a    & 3.111 \\
29     &  64.0345663 & -24.0669883  &    3.2215  & 14.1a  & 3.221 \\
29     &  64.0342617 & -24.0665167  &     -      & 14.1b  & 3.221 \\
29     &  64.0340392 & -24.0669400  &    3.2240  & 14.1d  & 3.221 \\
29     &  64.0461275 & -24.0768133  &     -      & 14.1f  & 3.221 \\
29     &  64.0340812 & -24.0664786  &    3.2215  & 14.1c  & 3.221 \\
30     &  64.0401537 & -24.0667703  &    3.2882  & 9.1b   & 3.292 \\
30     &  64.0451842 & -24.0723725  &     -      & 9.1a   & 3.292 \\
31     &  64.0436462 & -24.0590344  &    6.1452  & 2.1c   & 6.145 \\
31     &  64.0509446 & -24.0665642  &     -      & 2.1a   & 6.145 \\
31     &  64.0479096 & -24.0620906  &    6.1480  & 2.1b   & 6.145 \\
32     &  64.0377496 & -24.0607753  &    3.6065  & 6c     & 3.607 \\
32     &  64.0437075 & -24.0644186  &     -      & 6b     & 3.607 \\
32     &  64.0478788 & -24.0701956  &     -      & 6a     & 3.607 \\
33     &  64.0272008 & -24.0736031  &    3.9230  &  -     &  \\
33     &  64.0310679 & -24.0772083  &     -      &  -     &  \\
33     &  64.0312446 & -24.0770633  &     -      &  -     &  \\
34     &  64.0247004 & -24.0714867  &    2.5420  &  -     &  \\
34     &  64.0327150 & -24.0785378  &    2.5425  & 23b    & 2.542 \\
34     &  64.0357392 & -24.0799517  &     -      & 23a    & 2.542 \\
35     &  64.0309054 & -24.0837350  &    5.9729  & 33b    & 5.973 \\
35     &  64.0321100 & -24.0842606  &     -      & 33a    & 5.973 \\
36     &  64.0238171 & -24.0785033  &    3.9228  & 34c    & 3.923 \\
36     &  64.0277029 & -24.0826383  &     -      & 34b    & 3.923 \\
37     &  64.0222117 & -24.0773086  &    5.6380  & 35c    & 5.638 \\
37     &  64.0287462 & -24.0842464  &     -      & 35b    & 5.638 \\
37     &  64.0338062 & -24.0857294  &     -      & 35a    & 5.638 \\
38     &  64.0269388 & -24.0757711  &    2.3355  &  -     &  \\
38     &  64.0295250 & -24.0786211  &    2.3340  &  -     &  \\
38     &  64.0383083 & -24.0830053  &     -      &  -     &  \\
39     &  64.0272463 & -24.0682528  &    3.9680  & 17c    & 3.966 \\
39     &  64.0351867 & -24.0738864  &     -      & 17b    & 3.996 \\
39     &  64.0405671 & -24.0784167  &     -      & 17a    & 3.966 \\
40     &  64.0251204 & -24.0833403  &    2.2210  &  -     &  \\
40     &  64.0261767 & -24.0842808  &     -      &  -     &  \\
40     &  64.0309450 & -24.0867550  &     -      &  -     &  \\
\hline
 \end{tabular}  
 \end{minipage}
\end{table}

\setcounter{table}{0}
 \begin{table}
    \begin{minipage}{165mm}                                               
    \caption{cont.}
\begin{tabular}{|cccccc|}   
\hline
  ID   &       RA    &     DEC      & $z_{\rm MUSE}$ & ID$_2$ & $z_2$ \\
\hline  
41     &  64.0409938 & -24.0629858  &    2.2430  & 203c   & 2.245 \\
41     &  64.0411571 & -24.0630958  &     -      & 203b   & 2.245 \\
42     &  64.0366754 & -24.0661542  &    1.8270  & 202.2c & 2.091 \\
42     &  64.0369308 & -24.0663756  &     -      & 202.2b & 2.091  \\
43     &  64.0291179 & -24.0667792  &    1.9530  &  -     &  \\
43     &  64.0375096 & -24.0731747  &     -      &  -     &  \\
43     &  64.0386171 & -24.0739053  &     -      &  -     &  \\
44     &  64.0238937 & -24.0776206  &    1.8178  &  -     &  \\
44     &  64.0305688 & -24.0827056  &     -      &  -     &  \\
44     &  64.0325083 & -24.0837817  &     -      &  -     &  \\
45     &  64.0377446 & -24.0610592  &    4.1138  & 103c   & 4.115 \\
45     &  64.0429642 & -24.0639314  &     -      & 103b   & 4.115 \\
45     &  64.0482517 & -24.0709208  &     -      & 103a   & 4.115 \\
46     &  64.0258396 & -24.0722631  &    2.2800  &  -     &  \\
46     &  64.0329717 & -24.0770292  &     -      &  -     &  \\
46     &  64.0377117 & -24.0805028  &     -      &  -     &  \\
47     &  64.0362746 & -24.0606817  &    2.9220  & 107c   & 2.921 \\
47     &  64.0448367 & -24.0667225  &     -      & 107b   & 2.921 \\
47     &  64.0461029 & -24.0688289  &     -      & 107a   & 2.921 \\
48     &  64.0400717 & -24.0666814  &    3.2883  & 9.2b   & 3.292 \\
48     &  64.0454287 & -24.0725575  &     -      & 9.2a   & 3.292 \\
49     &  64.0343333 & -24.0630317  &    2.2810  & 8c     & 2.282 \\
49     &  64.0405508 & -24.0663597  &     -      & 8b     & 2.282 \\
49     &  64.0447008 & -24.0715192  &     -      & 8a     & 2.282 \\
50     &  64.0268658 & -24.0692417  &    3.2195  &  -     &  \\
50     &  64.0348675 & -24.0746150  &     -      &  -     &  \\
50     &  64.0397671 & -24.0788958  &     -      &  -     &  \\
51     &  64.0378375 & -24.0598592  &    4.1150  & 106c   & 4.116 \\
51     &  64.0459412 & -24.0658472  &     -      & 106b   & 4.116 \\
51     &  64.0477708 & -24.0686717  &     -      & 106a   & 4.116 \\
52     &  64.0239225 & -24.0750281  &    3.2910  & 204c   & 3.291 \\
52     &  64.0301767 & -24.0809514  &     -      & 204b   & 3.291 \\
52     &  64.0357446 & -24.0836114  &     -      & 204a   & 3.291 \\
53     &  64.0262950 & -24.0760453  &    2.9259  &  -     &  \\
53     &  64.0289225 & -24.0791542  &     -      &  -     &  \\
54     &  64.0267154 & -24.0747042  &    2.2300  &  -     &  \\
54     &  64.0302275 & -24.0786061  &     -      &  -     &  \\
54     &  64.0381763 & -24.0823422  &     -      &  -     &  \\
55     &  64.0263417 & -24.0687342  &    5.1060  &  -     &  \\
55     &  64.0349887 & -24.0746083  &     -      &  -     &  \\
55     &  64.0401338 & -24.0791333  &     -      &  -     &  \\
56     &  64.0366617 & -24.0633144  &    4.3000  & 101c   & 4.299 \\
56     &  64.0397071 & -24.0642781  &     -      & 101b   & 4.299 \\
56     &  64.0480804 & -24.0742222  &     -      & 101a   & 4.299 \\
57     &  64.0256154 & -24.0713078  &    4.5018  & 207c   & 4.502 \\
57     &  64.0331017 & -24.0760244  &     -      & 207b   & 4.502 \\
57     &  64.0393000 & -24.0813358  &     -      & 207a   & 4.502 \\
58\dag &  64.0235629 & -24.0731711  &    3.715   & 205c   & 3.715 \\
58     &  64.0311558 & -24.0803381  &    3.7153  & 205b   & 3.715 \\
58     &  64.0360404 & -24.0825781  &     -      & 205a   & 3.715 \\
59     &  64.0311383 & -24.0643317  &    4.0690  &  -     &  \\
59     &  64.0373113 & -24.0697069  &     -      & 104b   & 4.070 \\
59     &  64.0439900 & -24.0750847  &     -      & 104a   & 4.070 \\
60\dag &  64.0493708 & -24.0710067  &    6.145   & 112a   & 6.145 \\
60     &  64.0389625 & -24.0606803  &    6.1470  & 112c   & 6.145 \\
60     &  64.0433679 & -24.0629836  &     -      & 112b   & 6.145 \\
\hline
 \end{tabular}  
 \end{minipage}
\end{table}

\setcounter{table}{0}
 \begin{table}
    \begin{minipage}{165mm}                                               
    \caption{cont.}
\begin{tabular}{|cccccc|}   
\hline
  ID   &       RA    &     DEC      & $z_{\rm MUSE}$ & ID$_2$ & $z_2$ \\
\hline  
61\dag &  64.0508671 & -24.0664261  &    6.145   & 2.2a   & 6.145 \\
61     &  64.0434875 & -24.0589444  &    6.1480  & 2.2c   & 6.145 \\
61     &  64.0482517 & -24.0624331  &    6.1452  & 2.2b   & 6.145 \\
62     &  64.0335671 & -24.0650447  &    4.6090  & 108c   & 4.607 \\
62     &  64.0369179 & -24.0680367  &     -      &  -     &  \\
62\dag &  64.0366592 & -24.0680269  &    4.607   & 108b alt & 4.607 \\
62     &  64.0465858 & -24.0762047  &    4.6090  & 108a   & 4.607 \\
63     &  64.0333479 & -24.0674986  &    1.1472  & 201c   & 1.147 \\
63     &  64.0404400 & -24.0730881  &     -      & 201a   & 1.147 \\
64     &  64.0338125 & -24.0657972  &    4.0707  & 105c   & 4.071 \\
64     &  64.0360046 & -24.0678972  &     -      & 105b   & 4.071 \\
64     &  64.0464792 & -24.0767511  &     -      & 105a   & 4.071 \\
65     &  64.0392596 & -24.0698083  &    4.3000  & 110b   & 4.298 \\
65     &  64.0427467 & -24.0721897  &     -      & 110a   & 4.298 \\
66     &  64.0407246 & -24.0591806  &    5.0996  & 209c   & 5.100 \\
66     &  64.0468725 & -24.0638661  &     -      & 209b   & 5.100 \\
66     &  64.0496512 & -24.0680561  &     -      & 209a   & 5.0996 \\
67     &  64.0399400 & -24.0669239  &    5.9980  & 113b   & 5.995 \\
67     &  64.0459558 & -24.0740589  &     -      & 113a   & 5.995 \\
68     &  64.0364992 & -24.0622758  &    6.0664  & 102c   & 6.064 \\
68     &  64.0411162 & -24.0641319  &     -      & 102b   & 6.064 \\
68     &  64.0484808 & -24.0735939  &     -      & 102a   & 6.064 \\
69     &  64.0378317 & -24.0688725  &    2.9911  & 109b   & 2.989 \\
69     &  64.0438154 & -24.0737064  &     -      & 109a   & 2.989 \\
70     &  64.0289792 & -24.0755694  &    3.0750  &  -     &  \\
70     &  64.0296996 & -24.0762108  &     -      &  -     &  \\
71     &  64.0247200 & -24.0771819  &    4.5300  & 208c   & 4.530 \\
71     &  64.0277733 & -24.0806775  &     -      & 208b   & 4.530 \\
72     &  64.0308846 & -24.0837572  &    5.9730  &  -     &  \\
72     &  64.0320288 & -24.0842872  &     -      &  -     &  \\
73\dag &  64.0455825 & -24.0727003  &    3.2920  & 9.3a   & 3.29 \\
73\dag &  64.0400004 & -24.0666450  &    3.2920  & 9.3b   & 3.29 \\
74\dag &  64.0507508 & -24.0658519  &    6.1490  & 210a   & 6.149 \\
74\dag &  64.0472208 & -24.0611967  &    6.1490  & 210b   & 6.149 \\
74\dag &  64.0445658 & -24.0593369  &    6.1490  & 210c   & 6.149 \\
75\dag &  64.0414950 & -24.0626300  &    6.1490  & 210.4b & 6.149 \\
75\dag &  64.0401050 & -24.0618739  &    6.1490  & 210.4c & 6.149 \\
76\dag &  64.0458967 & -24.0602003  &    6.6290  & 211b   & 6.629 \\
76\dag &  64.0456046 & -24.0600436  &    6.6290  & 211c   & 6.629 \\

 \hline

 \end{tabular}
 \end{minipage}
\end{table}

\newpage

\end{appendix}

\end{document}